\documentclass[11pt]{article}
\usepackage{cite}
\usepackage{booktabs}
\usepackage[english]{babel}
\usepackage{amsmath, amssymb, amsbsy, amstext, amsthm, simplewick}
\usepackage{hyperref}
\usepackage{graphicx}
\usepackage{amsfonts}
\usepackage{upgreek}
\usepackage{framed}
\usepackage{pifont}
\usepackage{latexsym, mathrsfs}
\usepackage{array}
\usepackage{hyperref}
\usepackage{xspace}
\usepackage{longtable}
\usepackage{multirow}
\usepackage{cases}
\usepackage{empheq}
\usepackage[lofdepth,lotdepth]{subfig}
\usepackage[bottom]{footmisc}

\usepackage{braket}

\usepackage[load-configurations=astronomy, range-units=brackets, range-phrase=-, per-mode=reciprocal, mode=math]{siunitx}
\usepackage{threeparttable}
\usepackage{subfig}

\newcolumntype{C}[1]{>{\centering\let\newline\\\arraybackslash\hspace{0pt}}m{#1}}

\def\Mp{M_{\rm Pl}}

\def\mn{{\mu\nu}}
\newcommand{\cL}{{{\cal L}}}

\def\nn{{\nonumber}}
\newcommand{\bk}{{\boldsymbol{k}}}
\newcommand{\bp}{{\boldsymbol{p}}}
\newcommand{\bq}{{\boldsymbol{q}}}

\newcommand{\bx}{{\boldsymbol{x}}}
\newcommand{\bX}{{\boldsymbol{X}}}
\newcommand{\bT}{{\boldsymbol{T}}}
\newcommand{\bP}{{\boldsymbol{P}}}

\newcommand{\br}{{\boldsymbol{r}}}

\newcommand{\bv}{{\boldsymbol{v}}}

\newcommand{\bA}{{\boldsymbol{A}}}

\newcommand{\bn}{{\boldsymbol{n}}}
\newcommand{\ba}{{\boldsymbol{a}}}
\newcommand{\bb}{{\boldsymbol{b}}}

\newcommand{\bba}{\begin{array}}
\newcommand{\ea}{\end{array}}
\newcommand{\pa}[1]{\left(#1\right)}
\newcommand{\paq}[1]{\left[#1\right]}

\makeatletter
\newlength{\apb@width}
\newcommand{\autoparbox}[2][c]{\settowidth{\apb@width}{#2}\parbox[#1]{\apb@width}{#2}}

\makeatother

\numberwithin{equation}{section}

\def\beq{\begin{equation}}
\def\eeq{\end{equation}}

\def\bea{\begin{eqnarray}}
\def\eea{\end{eqnarray}}

\def\be{\begin{equation}}
\def\ee{\end{equation}}

\DeclareRobustCommand{\SkipTocEntry}[4]{}

\RequirePackage{color}

\usepackage{hhline}
\usepackage{array}
\newcolumntype{P}[1]{>{\centering\arraybackslash}p{#1}}
\usepackage[usenames,dvipsnames,svgnames,table]{xcolor} 
\usepackage{booktabs}
\usepackage{tikz}
\usetikzlibrary{calc,shadings,patterns,tikzmark}

\begin{document}
\setlength{\footnotesep}{\baselineskip}
\interfootnotelinepenalty=10000

\pagenumbering{roman}
\begin{titlepage}
\baselineskip=15.5pt \thispagestyle{empty}


\vspace{1cm}
\begin{center}
{\fontsize{18}{24}\selectfont  \bfseries  Conservative dynamics of binary systems to fourth Post-Newtonian order in the EFT approach II:  \\  [0.4cm] Renormalized Lagrangian} 
\end{center}
\begin{center}

\vskip8pt

{\fontsize{12}{18}\selectfont Stefano Foffa,$^{1}$ Rafael A. Porto,$^{2,3}$ Ira Rothstein,$^{4}$ and Riccardo Sturani\,$^{5}$} 
\end{center}

\begin{center}

\textsl{$^1$ \small D\'epartement de Physique Th\'eorique and Center for Astroparticle Physics,\\
Universit\'e de Geneve, 24 quai Ansermet, CH--1211 Geneve 4, Switzerland}  \vskip 6pt
\textsl{$^2$  \small Deutsches Elektronen-Synchrotron DESY,\\ Notkestra$\beta$e 85, D-22603 Hamburg, Germany}\vskip 6pt
\textsl{$^3$  \small The Abdus Salam International Center for Theoretical Physics,\\ Strada Costiera, 11, Trieste 34151, Italy}\vskip 6pt 
\textsl{$^4$ \small Department of Physics and Astronomy\\ Carnegie Mellon University, Pittsburgh, Pennsylvania 15213, USA} \vskip 6pt
\textsl{$^5$ \small International Institute of Physics (IIP),\\ Universidade Federal do Rio Grande do Norte (UFRN)\\
CP 1613, 59078-970 Natal-RN Brazil }
\end{center}

\hrule \vspace{0.3cm}
\noindent {\bf Abstract}\\[0.1cm]
We complete the derivation of the conservative dynamics of binary systems to fourth Post-Newtonian (4PN) order in the effective field theory (EFT) approach. We present a self-contained (ambiguity-free) computation of the renormalized  Lagrangian, entirely within the confines of the PN expansion. While we confirm the final results reported in the literature, we clarify~several issues regarding intermediate infrared (IR) and ultraviolet (UV) divergences, as well as the renormalization procedure. First, we properly identify the IR~and~UV singularities using (only) dimensional regularization and the method of regions, which are the pillars of the EFT formalism. This~requires a careful study of scaleless integrals in the potential region, as well as conservative contributions from radiation modes due to tail~effects. As expected by consistency, the UV divergences in the near region (due to the point-particle limit) can be absorbed into two counter-terms in the worldline effective theory.  The counter-terms can then be removed by field redefinitions, such that the renormalization scheme-dependence has no physical effect to 4PN~order. The remaining IR poles, which are spurious in nature, are unambiguously removed by implementing the zero-bin subtraction in the EFT approach. The procedure transforms the IR singularities into UV counter-parts. As~anticipated, the left-over UV poles explicitly cancel out against UV divergences in conservative terms from radiation-reaction, uniquely determining the gravitational potential. Similar artificial IR/UV poles, which are intimately linked to the split into regions, are manifest at lower orders. Starting at 4PN, both local- and nonlocal-in-time contributions from the radiation region enter in the conservative dynamics. Neither additional regulators nor ambiguity-parameters are introduced at any stage of the computations.\vspace{0.3cm} 
\hrule

\end{titlepage}

\thispagestyle{empty}
\setcounter{page}{2}
\tableofcontents

\clearpage
\pagenumbering{arabic}
\setcounter{page}{1}

\clearpage
\section{Introduction}
\label{sec:introduction}

\vskip 4pt

The detection of gravitational waves (GWs) by the LIGO/Virgo collaborations \cite{Abbott:2016blz} has initiated an unprecedented new epoch for explorations of the universe.  After the remarkable historical detections, GW science will soon turn into the study of the properties of the sources, addressing foundational questions in astrophysics, cosmology and particle physics~\cite{Buoreview, Porto:2016zng, Porto:2017lrn}. In~particular, binary systems of comparable masses or extreme-mass ratios are posed to become the leading probe to test gravitational dynamics and the physics of compact objects, such as black holes and neutron stars, under unique conditions. The number of events observed up to now demonstrates the feasibility of the direct detection of GWs over a large range of sources \cite{catalogue}. We expect several events per year once current detectors are running at designed sensitivity, and many more with future observatories. Precise theoretical templates are thus a compulsory ingredient for data analysis, and reliable physical interpretation of the signals, with the present and planned GW detectors. As a result, the two-body problem in gravity has become a very active area of research, relying on both numerical and analytical~methodologies~\cite{Buoreview}. A large portion of the GW signal is emitted during the inspiral phase, which in principle can be understood analytically using the Post-Newtonian (PN) formalism, where traditional methods in general relativity have a long and rich history, see e.g. \cite{Blanchet:2013haa,Schafer:2018kuf}. More recently, the Effective Field Theory (EFT) framework introduced in \cite{nrgr} has become a powerful new method to solve the two-body problem, successfully extending the knowledge of the binary's dynamics to high PN orders \cite{Foffa:2013qca, eftgrg20,Rothstein:2014sra, review, Levi:2018nxp}. The purpose of this paper is to continue the path towards precision gravitational waveforms, by completing the derivation of the gravitational potential for non-spinning compact objects to 4PN within the EFT approach, building upon the results reported in a companion paper \cite{Foffa:2019rdf} and elsewhere \cite{Foffa:2012rn,Foffa:2011np,Galley:2015kus,Foffa:2016rgu,Porto:2017shd,Porto:2017dgs}.\vskip 4pt

The Hamiltonian (and Lagrangian) for the conservative dynamics at 4PN was first reported~in \cite{Jaranowski:2013lca,Bini,Damour:2014jta,Jaranowski:2015lha,Damour:2016abl} and \cite{Bernard:2015njp,Bernard:2016wrg} in the ADM~\cite{Schafer:2018kuf} and `Fokker-action'~\cite{Blanchet:2013haa} approaches, respectively. Yet,~the introduction of `ambiguity-parameters' due to the presence of infrared (IR) divergences, in addition to the ultraviolet (UV) ones, together with further claims for more ambiguities to address a discrepancy between the two first independent derivations \cite{Damour:2016abl,Bernard:2015njp,Bernard:2016wrg}, appeared to signal a breakdown of the split into regions at 4PN order. As~a~consequence, the ambiguities were {\it resolved} originally by relying on information outside of the PN framework \cite{Bini,Damour:2014jta,Bernard:2016wrg}. Later on, the rederivation in \cite{Bernard:2017bvn,Marchand:2017pir} of the conservative contribution in radiation-reaction due to the tail effect using dimensional regularization (dim. reg.), confirming the result obtained in \cite{Foffa:2011np,Galley:2015kus} within the EFT approach, provided the last ingredient to fix the two ambiguities introduced in~\cite{Bernard:2015njp,Bernard:2016wrg}. This completed the calculation within the Fokker-action formalism, without the need of extra matching conditions. 
The one ambiguity-parameter in the ADM approach \cite{Jaranowski:2013lca,Damour:2014jta,Jaranowski:2015lha} has, thus far, only been obtained incorporating results from gravitational~self-force~calculations \cite{Bini}.
\vskip 4pt Even though, in practice, the ambiguities introduced in \cite{Jaranowski:2013lca,Damour:2014jta,Bernard:2015njp,Bernard:2016wrg,Jaranowski:2015lha,Damour:2016abl} were determined and the complete result reported, both the derivations in the ADM \cite{Jaranowski:2013lca, Jaranowski:2015lha} and Fokker-action \cite{Bernard:2017bvn,Marchand:2017pir} formalisms left room open for further improvement and clarifications. In~particular, with regards to the handling of IR divergences (which led the introduction of ambiguity-parameters in the first place) and the apparent reliance on an extra regulator beyond dim. reg.\footnote{A joined (``$\epsilon B$") dimensional and analytic regularization is used in the ADM formalism [see Appendix~A in \cite{Jaranowski:2015lha}] while, similarly, a combined (``$\epsilon\eta$") regularization is implemented in the Fokker-action approach [see e.g. the paragraphs after Eq. (3.6) in \cite{Bernard:2017bvn} and after Eqs. (2.5)-(2.6) in \cite{Marchand:2017pir}].} Moreover, while the renormalization procedures presented elsewhere led to the correct result,\footnote{For instance, a (short-distance) worldline shift is implemented in \cite{Bernard:2015njp} (see e.g. their Appendix~C). However, it includes both (long-distance) IR poles from the potential region and UV poles from tail terms  {\it combined}, together with~their associated length scales.} a more systematic removal of IR/UV divergences will be needed when physical logarithms in the near zone first appear at higher PN orders (due to finite-size effects). We~address all of these issues in the present paper, by providing an ambiguity-free and systematic derivation of the renormalized Lagrangian in dim.~reg., all within the confines of the PN expansion, which can be naturally extended to all orders.\vskip 4pt

As it was discussed in \cite{Galley:2015kus,Porto:2017dgs,Porto:2017shd}, the EFT formalism clearly illustrates the origin of the apparent ambiguities due to the split into regions, while already providing the unambiguous contributions from the tail effect to the effective Lagrangian. However, the computation of the local-in-time near zone regularized Lagrangian in the EFT approach was (until now) pending, with intermediate results at orders $G, G^2$ and $G^5$ presented elsewhere \cite{Foffa:2012rn,Foffa:2016rgu} (see also \cite{Damour:2017ced}).\footnote{The static $G^6$ potential at 5PN was recently computed in \cite{Foffa:2019hrb} (see also \cite{Blumlein:2019zku}).} In~a~companion paper \cite{Foffa:2019rdf}, the remaining $G^3$ and $G^4$ contributions are reported, using dim. reg. to handle the divergences. We point the reader to \cite{Foffa:2019rdf} for a thorough derivation of the relevant Feynman diagrams which contribute to this order. While, in the near region, the singular terms in the limit $d \to 3$ were identified in \cite{Foffa:2019rdf}, the distinction between IR and UV poles was not addressed. This is crucial for the proper renormalization of the effective theory. One of the goals of this paper is therefore to perform a careful analysis of divergent integrals in dim. reg., identifying   the type of near and far zone singularities. As we shall see, scaleless (self-energy) integrals as well as conservative radiation-reaction terms play a key~role.\vskip 4pt After we isolate the coefficients of the IR and UV poles, we perform the systematic renormalization of the effective theory. By~identifying the IR/UV singularities, the elimination of UV poles in the potential region can be performed without knowledge of contributions from radiation modes, as expected. These UV poles can be absorbed into counter-terms in the point-particle action, which in turn can be removed by field-redefinitions, as emphasized in \cite{nrgr}. Therefore, the renormalization scheme-dependence has no physical effect to 4PN~order and, for simplicity, we will choose a minimal-subtraction (MS) scheme. Once the UV poles are renormalized away, the remaining IR divergences, arising in the near zone due to an overlap (double-counting) between regions of integration, are handled by the zero-bin subtraction \cite{zb} applied to the EFT approach \cite{Porto:2017dgs,Porto:2017shd}. The procedure unambiguously removes the IR poles, transforming them into UV counter-parts. The left-over poles cancel out against divergences arising in conservative radiation-reaction terms, uniquely fixing the~gravitational~potential at~4PN order, as it was emphasized~in~\cite{Galley:2015kus,Porto:2017dgs,Porto:2017shd}.\footnote{The explicit cancelation between spurious near/far zone divergences is not manifest in the ambiguity-free derivation within the Fokker-action approach \cite{Bernard:2017bvn,Marchand:2017pir}, which instead relies on an additional worldline redefinition to remove the remaining IR/UV poles~\cite{Bernard:2015njp,Bernard:2016wrg}.}\vskip 4pt As we shall see, the link between IR/UV divergences appears already at lower orders, as it is required by consistency of the split into regions, albeit with contributions which are proportional to conserved currents or vanish on-shell. The situation changes at 4PN, where the cancellation of spurious divergences after the subtraction of the zero-bin leaves behind physical contributions to the effective action \cite{Porto:2017dgs}. On the one hand, it includes a term which mirrors the celebrated factor of $5/6$ in the Lamb shift in QED \cite{Galley:2015kus,Porto:2017shd}. On the other hand, there is also a nonlocal (in time) contribution which resembles instead the Bethe logarithm \cite{Galley:2015kus,Porto:2017shd}. The~final form of the effective action turns out to be equivalent to the one reported in \cite{Damour:2014jta,Marchand:2017pir}, leading to the same expression for physical observables, such as the binding energy and periastron advance, following the careful treatment of the non-local term discussed in \cite{Damour:2016abl, Bernard:2016wrg}. Our derivation thus supports the validity of the 4PN results, confirmed by three independent methodologies. At the same time, the computation within the EFT approach improves on the previous computations in \cite{Jaranowski:2013lca,Damour:2014jta,Jaranowski:2015lha,Bernard:2015njp,Damour:2016abl,Bernard:2016wrg,Bernard:2017bvn,Marchand:2017pir}. Most notably, as anticipated~in~\cite{Galley:2015kus,Porto:2017dgs,Porto:2017shd}, neither ambiguity-parameters nor additional regulators are required at any stage of our derivation.\vskip 4pt

The present paper is divided as follows. Next, in sec.~\ref{sec:eft}, we review the EFT formalism, with emphasis on the method of regions and IR/UV divergences with potential and radiation~modes. In~sec.~\ref{sec:renorm}, we isolate the intermediate IR and UV poles in the computation of the near zone conservative dynamics, and discuss the renormalization procedure to remove the UV divergences through counter-terms. As we shall see, one of the counter-terms is already fixed at 3PN~order and readily removes most of the 4PN divergences. The remaining (few) UV poles are taken care of by a second counter-term, which starts at 4PN. In sec. \ref{sec:cancel} we discuss the subtraction of the zero-bin, which removes the IR singularities from the near zone. We demonstrate the explicit cancellation of the left-over UV poles, uniquely determined from this procedure, against UV divergences in conservative radiation-reaction effects from the far zone. In sec. \ref{sec:renormtot}, after removing unphysical long- and short-distance logarithms, we present the final form of the renormalized Lagrangian, including local- and nonlocal-in-time contributions. We conclude in sec. \ref{sec:disc}, with comments on the origin of the spurious IR/UV divergences. Details are relegated to the appendices.  
\subsubsection*{Conventions}
Throughout this paper we use the following notational conventions. For spacetime variables, which depend on the proper time, $\tau_a$, we use $v^\alpha_a(\tau_a) \equiv \dot x^\alpha_a (\tau_a) \equiv \frac{d x_a^\alpha(\tau_a)}{d\tau_a}$, $\dot v^\alpha_a(\tau_a) = \frac{d v_a^\alpha(\tau_a)}{d\tau_a}$, $a^\alpha_a (\tau_a) \equiv \frac{D v_a^\alpha(\tau_a)}{D\tau_a} \equiv \dot v^\alpha_a(\tau_a) + \Gamma^\alpha_{\mu\nu} v_a^\mu(\tau_a) v_a^\nu(\tau_a)$, where $x_a^\mu(\tau_a)$ describes the particle's worldline $(a=1,2)$. For the three dimensional variables, which depend instead on the coordinate time, $t$, we use $\bv_a(t) \equiv \dot \bx_a(t) \equiv \frac{d\bx_a(t)}{dt}$,  $\ba_a(t) \equiv \dot \bv_a(t), \bb_{a}(t) \equiv \dot \ba_a(t)$, $\br(t) \equiv \bx_1(t)-\bx_2(t)$, $\bv(t) \equiv\dot{\br}(t)=\bv_1(t)-\bv_2(t)$, $\ba(t) \equiv \dot \bv(t)=\ba_1(t)- \ba_2(t)$, and $\bb(t) \equiv\dot \ba(t) = \bb_1(t)- \bb_2(t)$.

\section{Effective field theory approach}\label{sec:eft}

We summarize first the basic elements of the EFT framework put forward in~\cite{nrgr}, and further developed in \cite{dis1,nrgrs,prl,dis2,Porto:2007tt,Gilmore:2008gq,nrgrss,nrgrs2,nrgrso,andirad,srad1,srad2,Foffa:2011ub, andi,Goldberger:2012kf,Levi:2016ofk}, with emphasis on the aspects discussed in \cite{Galley:2015kus,Porto:2017dgs,Porto:2017shd}. For reviews of the EFT formalism applied to the binary inspiral problem see \cite{Goldberger:2007hy, Porto:2007pw,Foffa:2013qca,eftgrg20,Rothstein:2014sra,review,Levi:2018nxp}.  
\subsection{Point-particle action}

When compact bodies are probed on scales larger than their typical sizes, it is justified to write an effective theory describing a collection of worldlines (around e.g. the center-of-mass of each particle) interacting with the gravitational field.
The dynamics is described by an effective action:
\beq
\label{eq:spp}
S_{\rm pp}[x^\alpha_a(\tau_a)] = \sum_a \int d \tau_a \left(-m_a +  \sum_i c_i {\cal O}_i\big[x_a^\alpha(\tau_a),v_a^\alpha(\tau_a), \cdots ; g_{\mu\nu}, \partial_\beta g_{\mu\nu}, \cdots\big]\right)\,.
\eeq
Following the jargon of quantum field theory, we often use the term {\it operators} to denote the ${\cal O}_i$'s. These operators are invariant under the relevant symmetries, namely diffeomorphism and worldline reparameterizations, once on-shell conditions are imposed for the metric and matter~fields.\vskip 4pt

Since we must ultimately choose a gauge when performing intermediate calculations, the need of operators which are not manifestly invariant off-shell will become relevant when discussing the renormalization of the theory. For example,
the operator
\beq
\label{eqOau}
{\cal O}_{a\dot v} = g_{\mu\nu} a^\mu \dot v^\nu\,
\eeq
is allowed by symmetries. Notice, this extra term vanishes on-shell for non-spinning bodies, due to geodesic motion (but see footnote~\ref{footspin} in sec.~\ref{sec:disc}). Nevertheless, operators which are zero on-shell play an important role in removing divergences which turn out to be proportional to lower-order the equations of motion. At the end of the day, they can be removed by field redefinitions.\footnote{Any term in the action proportional to the leading order equations of motion $D\varphi=0$, i.e. $F[\varphi]D\varphi$, where $F[\varphi]$ is some polynomial in the fields and their derivatives, can be removed by a transformation $\varphi \rightarrow \varphi-F[\varphi]$, where $(\bx_a,g_{\mu\nu})\in \{\varphi\}$ in our case.}\vskip 4pt

Other type of {\it kinematic} operators, such as $a^\mu a_\mu\,,$ may also be added. However, in our case this operator is a `double-zero' for non-spinning bodies, and therefore it can be ignored. (See \cite{Galley:2010es,Forgacs:2012qt,Galley:2012gv} for a discussion of acceleration-dependent operators to describe finite size effects in electrodynamics.) Another class of operators, which also vanish on-shell but depend on the Riemann tensor \cite{nrgr}, are
\beq
{\cal O}_R \equiv R^\alpha_\alpha,\,\, {\rm or}\,\, {\cal O}_{V} \equiv R_{\mu\nu}v^\mu v^\nu\,,\label{eqRuu}
\eeq
with $R_{\mu\nu}$ the Ricci tensor. The operators in \eqref{eqRuu} were introduced in \cite{nrgr} to regularize the one-point function in the static limit, while the operator in \eqref{eqOau} (which is only invariant on-shell) enters for non-static sources. While the coefficients of these particular operators are not linked with physical effects, they are required to consistently remove the UV poles in harmonic gauge.\vskip 4pt

There are, of course, physical parameters associated to finite-size terms. For instance, tidal deformations are described (at leading order) by the finite-size operator, ${\cal O}_{E^2} \equiv E_{\mu\nu} E^{\mu\nu}$, with $E_{\mu\nu}$ the electric component of the Weyl tensor. Its coefficient is often called the (electric) tidal `Love number'. (A similar term may be written in terms of the magnetic component.) This operator is not relevant until 5PN order, and therefore does not play a role in our discussion.

\subsection{Potential region}

The EFT described by \eqref{eq:spp} has no reference to the PN expansion. However, once the compact bodies belong to a non-relativistic bound state, with typical separation $r$, it is useful, for the purposes of manifest power counting, to decompose the metric field into ``potential" ($H_\mn$) and ``radiation" ($\bar h_\mn$) modes, varying on scales $(k^0,|\bk|)_{\rm pot} \simeq (v/r,1/r)$ and $(k^0,|\bk|)_{\rm rad} \simeq (v/r,v/r)$ respectively, with $v$ the relative velocity \cite{nrgr}. Notice that the distinction is only meaningful for $v \ll 1$, such that the perturbative expansion in the ratio of relevant scales is ultimately organized in powers of $v$. By~solving~for (or {\it integrating-out}) the quasi-instantaneous modes order by order, one can compute the contribution to the conservative sector from potential modes. The~calculation of the relevant Feynman diagrams at 4PN order was performed in \cite{Foffa:2019rdf}, which we encourage the reader to consult for further details (see appendix~\ref{sec:app0} for a brief summary). Because of the split into regions, spurious divergences develop in the intermediate steps. We~comment below on their origin and how they are handled by the EFT approach. 

\subsubsection*{UV Divergences}

\begin{figure}[t!]
\centerline{\scalebox{1.3}{\includegraphics{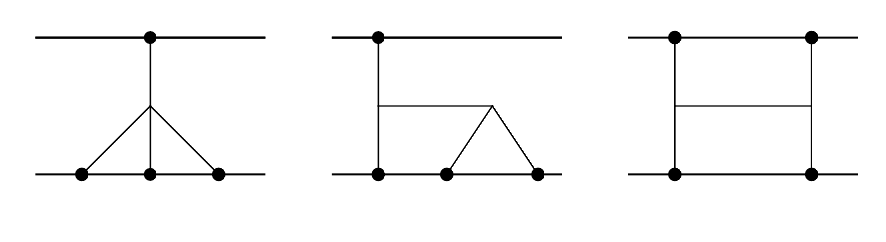}}}
\caption[1]{Topologies contribution at ${\cal O}(G^3)$ to the Lagrangian. The first two diagrams are UV divergent, starting at 3PN, while the third diagram is finite at this order. All of these topologies are divergent at 4PN. The first diagram is UV while the third is IR divergent. The second one has both, IR and UV, poles.} \label{fig:3pnUV}
\end{figure}

As it is well known, when working in a non-linear classical theory such as general relativity, point-like sources introduce UV singularities. In dim. reg., with the number of space dimensions being $d=3+\epsilon$, the logarithmically UV divergent integrals lead to $1/\epsilon_{\rm UV}$ poles, as $\epsilon_{\rm UV} \to 0^-$. On the other hand, power law divergences in dim. reg. are set to zero. For instance, at 3PN order, the first two diagrams in Fig.~\ref{fig:3pnUV} are logarithmically UV divergent.  As it was discussed in \cite{nrgr}, these classical divergences are treated as in standard quantum field theory, by writing the bare effective action in terms of a counter-term and renormalized parameters $c_{\alpha \, \rm bare}= c_{\alpha,\, \rm c.t.}+ c_{\alpha,\, \rm ren}(\mu)$, and choosing $c_{\alpha,\, {\rm c.t.}}$ to cancel the poles to render the result finite.
A~renormalization scale, $\mu$, is introduced in dim. reg. to account for the change in dimensions in Newton's constant, e.g. \cite{andirad}, \beq \label{eq:gd} G_d = \mu^{3-d} G\,.\eeq  This introduces factors of $\log \mu$'s in the $d \to 3$ limit when poles are present, as well as Euler's constant, $\gamma_E$, from the expansion of the associated $\Gamma$-functions.  For ease of notation in the near zone computations we will often use the combination 
\beq
\label{barmu} \bar\mu \equiv \mu\, \sqrt{4\pi} e^{\gamma_E/2}\,,
\eeq
which recurrently appears in the regularization in the potential region. The $\mu$-dependence is absorbed into the renormalized parameters, rendering the results independent on the choice of renormalization scale. See sec.~\ref{sec:renormtot} for more details.
\vskip 4pt
The subtraction is of course naively ambiguous, as one is free to add any finite amount to the renormalized coefficients. For the cases when the divergence is `physical', in the sense that it does not vanish on-shell, the subtraction constant is fixed by a {\it matching} procedure. Typically, this is performed by calculating response functions in external backgrounds, within the overlapping realm of validity of the full and effective theory, e.g. \cite{Damour:2009vw,Binnington:2009bb,Kol:2011vg}. For our case, since the operators which will be needed to absorb all of the UV divergences in the near region can be removed by field redefinitions, the scheme-dependence in the choice of counter-terms does not have physical effect, resulting in unambiguous results.

\subsubsection*{IR singularities}

\begin{figure}[!t]
\centerline{\scalebox{1.4}{\includegraphics{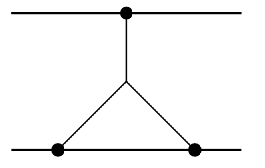}}}
\caption[1]{The first non-linear topology leading to IR divergences away from the static limit. The spurious poles occur when the propagators are Taylor expanded in powers of $p_0/|\bp|$.} \label{fig:Y}
\end{figure}

Naively, IR divergences can also show up in the computation of the near zone potential. However, since the binding is generated from modes whose wavelength is cut-off by the orbital scale, any IR divergence in the near region must be spurious in nature. Ultimately, any such poles are due to the fact that we do not impose a hard cut-off on the potential modes, to avoid an explicit mutilation of diffeomorphism invariance. This is also one of the main reasons to implement dim. reg. in the intermediate calculations. As a consequence, when we perform momentum integrals in $d=3+\epsilon$ over there full range of scales we may encounter, not only UV poles, but also IR singularities as $\epsilon_{\rm IR} \to 0^+$. For instance, IR divergences appear already in the diagram with the topology in Fig.~\ref{fig:Y} at ${\cal O}(G^2)$, with velocity corrections to the propagators of potential modes.\footnote{This diagram first enter at 2PN order \cite{Gilmore:2008gq}. Further velocity corrections produce extra IR poles at higher PN orders. Other sources of IR (and UV) divergences appear in topologies at higher order in $G$, as in Fig.~\ref{fig:3pnUV}.}
Let us emphasize that these are entirely a byproduct of the splitting into regions. The IR-sensitivity is induced by the quasi-instantaneous expansion of Green's functions in powers of $p^0/|\bp|$, and therefore are not present for un-expanded propagators. See sec.~\ref{sec:disc} for more details.\vskip 4pt
Unlike UV divergences, IR singularities are due to an overlap between different regions of integration. This double counting is a well known phenomenon in EFTs, and there is a systematic framework, known as the zero-bin subtraction \cite{zb}, which was developed precisely to handle this issue.  In the present context,  the zero-bin subtraction was discussed in \cite{Porto:2017dgs,Porto:2017shd}. While, as we shall see, in dim. reg. the zero-bin subtraction to 4PN order amounts to replacing $\epsilon_{\rm IR} \to \epsilon_{\rm UV}$ through scaleless integrals, the procedure is in principle independent of the regulator \cite{zb}. Therefore, it can be also incorporated in other formalisms, in particular to remove the ambiguity-parameter(s) introduced in the derivations presented in \cite{Damour:2014jta,Bernard:2015njp}. As we demonstrate explicitly, the resulting UV poles after the subtraction of the zero-bin cancel out against conservative contributions from the radiation region, uniquely fixing the gravitational potential.

\subsection{Radiation region} 

After the potential modes are integrated out, the long-distance effective action for the binary system, now treated as a point-like object, is written in the form of a multipole expansion. The~effective~action takes the form (e.g. around the center-of-mass, $\bX$, of the binary)
\bea
\label{eq:srad}
S_{\rm rad} &=& \int d \tau \left[ -P^\mu(\tau) V^\nu (\tau) \left(1 + \bX^i \nabla_i\right)\bar h_{\mu\nu}(\tau,\bX(\tau)) - \frac{1}{2} \bar\omega^{\alpha\beta}_\mu (\tau,\bX) J_{\alpha\beta}(\tau) V^\mu (\tau) \right. \\ &&+ \left. 
\sum_{\ell=2} \left( \frac{1}{\ell!} I^L_{\rm STF} (\tau) \nabla_{L-2} \bar E_{i_{\ell-1}i_\ell}(\tau,\bX(\tau)) - \frac{2\ell}{(2\ell+1)!}J_{\rm STF} ^L(\tau) \nabla_{L-2} \bar B_{i_{\ell-1}i_\ell}(\tau,\bX(\tau))\right)\right]\, . \nn
\eea 
We use the shortened notation $L \equiv \left\{i_1\ldots i_\ell\right\}$, such that $\bx^L\equiv \bx^{i_1}\cdots \bx^{i_\ell}$. The four-momentum of the binary system is given by $P^\mu(\tau) = M V^\mu(\tau)$, with $V^\mu$ its four-velocity and $M$ the binding mass-energy,  whereas $\omega_\mu^{\alpha\beta}$ are the Ricci rotation coefficients which couple to the angular-momentum tensor $J_{\alpha\beta}$. The {\it source} multipole moments, $(I^L_{\rm STF}(\tau),J^L_{\rm STF} (\tau))$, are $SO(3)$ symmetric and trace-free (STF) tensors.  The bar indicates that geometric quantity ought to be evaluated on the radiation field $\bar h_{\mu\nu}$.\vskip 4pt

The multipoles in \eqref{eq:srad} can be written in terms of moments of the stress-energy tensor using relations that rely upon the use of on-shell conservation laws \cite{andi}. This can be illustrated with the quadrupole coupling. For simplicity, let us assume the binary is at the origin at rest, such that $\bX=\dot \bX = {\bf 0}$. Hence, the leading term in the multipole expansion takes the form (in $d$ dimensions)
\bea
\label{eq:con}
\int dt \left(\int d^d\bx\, {\cal T}^{ij}(t,\bx)\right) \bar h_{ij}(t,{\bf 0}) &=& \int dt \, \left( \int d^d \bx \, {\cal T}^{00}(t,{\bx})\bx^i\bx^j\right) \frac{1}{2} \partial_t^2 \bar h_{ij}(t,{\bf 0})   \\ &&-  \int dt \, \left(2\int d^d \bx \, \partial_\mu {\cal T}^{\mu i}(t,{\bx})\bx^j \right) \bar h_{ij}(t,{\bf 0}) \nn \\ &&-   \int dt \, \left(\frac{1}{2} \int d^d \bx \, \partial_\mu\partial_\nu {\cal T}^{\mu\nu}(t,{\bx})\bx^i\bx^j\right) \bar h_{ij}(t,{\bf 0})\,. \nn
\eea
We can then rewrite the first term as (after including other components of the metric tensor)
\beq
\int dt\, I_0^{ij}(t) E_{ij}\,,
\eeq
with 
\beq
\label{eq:LOI}
I_0^{ij}(t) = \int d^d \bx \, {\cal T}^{00}(t,\bx) \bx^i\bx^j\,,
\eeq
at leading order in the multipole expansion. To obtain the expression in terms of the constituents of the binary we must perform a matching computation \cite{andirad}, see appendix \ref{sec:appA}. The last two terms in \eqref{eq:con} vanish on-shell, and therefore may be discarded for the derivation of physical observables. However, as we shall see, they will play an important role to cancel out many of the spurious divergences arising in the potential region. 

\subsection{Radiation-reaction}
\begin{figure}[!t]
\centerline{\scalebox{0.35}{\includegraphics{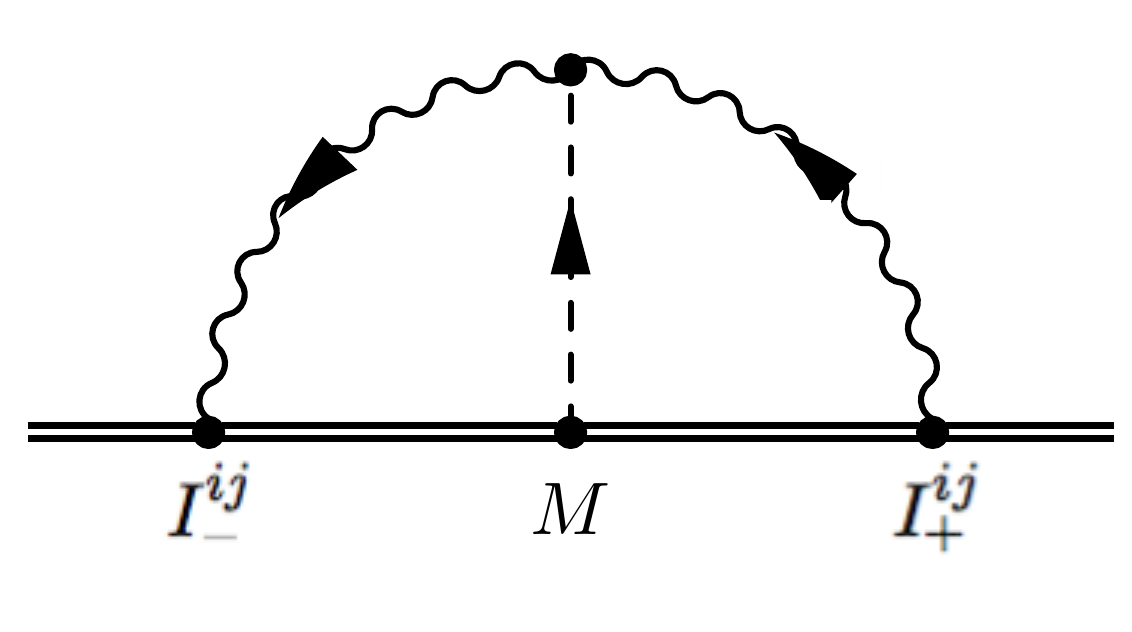}}}
\caption[1]{Feynman diagram for the contribution to the radiation-reaction force due the tail effect. See \cite{Galley:2015kus} for more details.} \label{nltail}
\end{figure}
Armed with a long-distance effective theory we can readily compute the radiation-reaction force produced by GW emission. As it was shown in \cite{Galley:2015kus} in the context of the EFT approach, there are both a dissipative {\it and} a conservative contribution due to GW radiation scattering off of the background geometry. To compute radiation-reaction effects we use the in-in formalism adapted to a classical setting (see~\cite{Galley:2012hx,Galley:2014wla} for further details). At 4PN we have the tail diagram in Fig.~\ref{nltail}, which is UV divergent. The result in dim. reg. reads (omitting the STF label) \cite{Galley:2015kus}
\begin{align}
S_{\rm 4PN}^{\rm tail}[\bx_a^\pm]  =   \, \frac{2 G^2 M}{5} \int_{-\infty}^\infty \frac{d\omega}{2 \pi} \,  \omega^6 \, I_-^{ij}(-\omega) I_+^{ij}(\omega) \bigg[ & -\frac{1}{ \epsilon_{\rm UV}} - \gamma_E + \log\pi -  \log \frac{\omega^2}{\mu^2} + \frac{41}{30} + i \pi \, \text{sign}(\omega) \bigg]\,,
\label{eq:RRnl}
\end{align}
where 
\begin{align}
	I_{-}^{ij} (t) & \equiv I^{ij} (t, \bx_a^{(1)}) - I^{ij} (t, \bx_a^{(2)} )  =  \sum_a m_a \left(\bx_{a-}^{i} \bx_{a+}^{j} +  \bx_{a+}^{i} \bx_{a-}^{j} - \frac{2}{3}\, \delta^{ij} \bx_{a-} \! \cdot \! \bx_{a+}\right)  + {\cal O} (\bx_{a -}^3 ) , \nn \\
	I_{+}^{ij} (t) & \equiv \frac{1}{2} \Big( I^{ij} (t, \bx_a^{(1)}) + I^{ij} (t, \bx_a^{(2)} ) \Big) = \sum_a m_a \left( \bx_{a+}^i \bx_{a+}^j - \frac{1}{3} \delta^{ij} \bx_{a+}^2 \right) + {\cal O} (\bx_{a -}^2 ).
\end{align}
We will be concerned with the conservative sector in this paper, which is symmetric under $\omega \to -\omega$.\footnote{The dissipative term, $i\pi \,\text{sign}(\omega)$, reproduces the well-known tail correction to the energy flux, e.g. \cite{andirad}.}
The correction to the equation of motion follows from 
\begin{equation}
\label{eq:St}
\left[ \frac{\delta S_{\rm tail}[\bx^{a\pm}]}{\delta \bx_{a-} } \right]_{\rm PL} = 0\,, 
\end{equation}
where the ``PL'' subscript indicates the ``physical limit'' for which the ``$-$'' variables vanish and the ``$+$'' variables are set to their physical values \cite{Galley:2012hx}. In practice, this means that the result is half of that in \eqref{eq:RRnl} in ``standard'' variables, where the derivative is then allowed to hit both multipoles. From \eqref{eq:St} we obtain the radiation-reaction acceleration. At 4PN order, \eqref{eq:RRnl} provides an essential (and uniquely determined) contribution to the binary's dynamics similar to the Lamb~shift \cite{Galley:2015kus,Porto:2017dgs,Porto:2017shd}.
\vskip 4pt

Notice that in the computation of the tail effect we have ignored multipole moments which are conserved, or terms that vanish on-shell. For instance, the linear and angular momentum, as well as the extra pieces that appear following the manipulations in \eqref{eq:con}, have not been included. Yet, inserted in a diagram similar to Fig.~\ref{nltail}, these terms also produce UV divergent integrals similar to the quadrupole contributions in \eqref{eq:RRnl}. As we shall demonstrate explicitly in sec.~\ref{sec:cancel}, the resulting UV poles play a key role in cancelling the associated spurious singularities which develop in the potential region. As we shall see, the cancellation starts at ${\cal O}(G^2)$, and already at 2PN order, which is required for the consistency of the near/far zone descriptions. 


\section{Renormalization of near zone UV divergences}\label{sec:renorm}
To renormalize the effective theory, the UV poles in the near zone must be absorbed into the counter-terms in \eqref{eq:spp}. We show here how this is implemented to 4PN order. First, we will show how to identify (when IR singularities are present) the coefficient of both UV and IR poles. This requires incorporating self-energy (scaleless) integrals. The renormalization of divergences in the potential region then proceeds without inputing information from long-distance modes, as expected. In section~\ref{sec:cancel} we discuss the subtraction of the remaining IR singularities.

\subsection{Potential region IR/UV poles}\label{secnearz}

\subsubsection*{2PN order}

\begin{figure}[!t]
\centerline{\scalebox{0.4}{\includegraphics{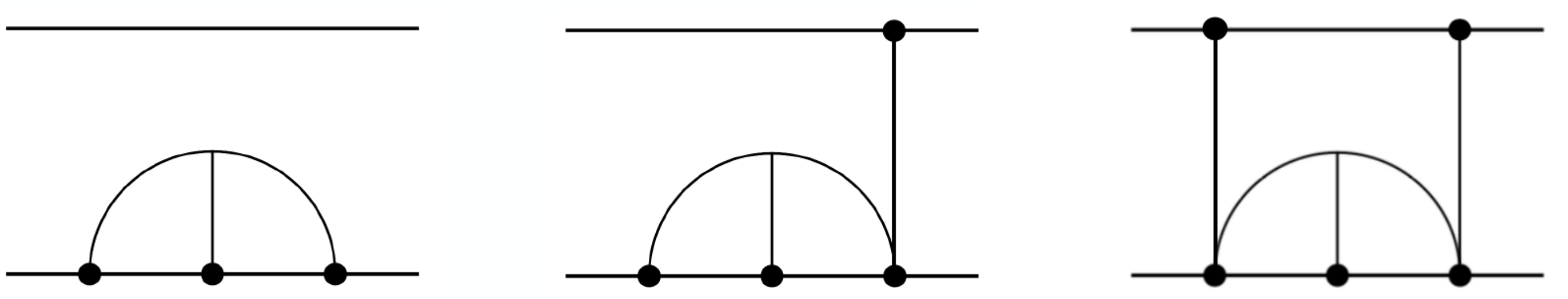}}}
\caption[1]{Self-energy diagrams leading to scaleless integrals with logarithmic IR/UV poles at $G^2, G^3$ and $G^4$, respectively. Unlike the first, which does not depend on the mass of the companion, the other diagrams represent self-energy corrections to the gravitational~interaction.} \label{fig:Y3}
\end{figure}

There are no logarithmic divergences at 2PN order. That is the case, provided we take the mass of the compact object to be time-independent. However, it is easy to show that, for the sake of argument, allowing for a non-trivial time-dependence ($\dot m_a(t) \neq 0$) the resulting 2PN effective Lagrangian develops logarithmic IR poles. These are due to the diagram in Fig.~\ref{fig:Y},
and take the form
\beq
\cL^{\rm IR \, (near)}_{\rm 2PN} = \frac{1}{\epsilon_{\rm IR}} G^2\left(\dot m_1^2m_2+2 \dot m_1\dot m_2 m_1\right) + \left(1\leftrightarrow 2\right)\,.
\eeq 
The form of this divergence, at this stage, is not particularly illuminating. However, as we emphasized, to properly identify the coefficient of the IR poles in the near region we must also add self-energy contribution. At 2PN, this is represented by the first diagram in Fig.~\ref{fig:Y3}.
The result is given by
\beq
\cL^{\rm self}_{\rm 2PN} = \left(\frac{1}{\epsilon_{\rm IR}}-\frac{1}{\epsilon_{\rm UV}}\right) G^2 \dot m_1^2m_1 +\left(1\leftrightarrow 2\right)\,.
\eeq
Adding the pieces together we end up with both IR and UV poles:
 \beq
\label{eq:div2pn}
\cL^{\rm IR/UV \, (near+self)}_{\rm 2PN} = \frac{1}{\epsilon_{\rm IR}} G^2M_0 \dot M_0^2
-\frac{1}{\epsilon_{\rm UV}}G^2 \pa{m_1 \dot{m}_1^2 +m_2\dot m_2^2}\,.
\eeq
The alert reader will immediately realize that the IR singularity now has the form of a monopole-radiation coupling proportional to $\dot M_0$, with $M_0=m_1+m_2$ at this order.  As we have anticipated, this is consistent with the fact that this IR divergence will be associated to a contribution from the radiation modes. (The coupling to the monopole in the far zone, entering in  an expression similar to \eqref{eq:div2pn}, will be relevant later on.) The near region UV divergence at this order may be absorbed into a mass renormalization. This will not be the case at higher PN orders. In~what follows we will set $\dot m_a=0$, also for the divergent terms, which we kept here solely for pedagogical~reasons. 

\subsubsection*{3PN order}

At this order both IR and UV divergences are present.\footnote{In the derivation in \cite{Foffa:2011ub}, the divergences are computed in dim. reg. as poles in $(d-3)$, regardless of their IR or UV nature. Since, as we shall see, there is no physical contribution from radiation-reaction at this order, this does not affect the 3PN Lagrangian. However, as we emphasized, the distinction is essential at 4PN order.} The IR poles, once again, arise from the topology in Fig.~\ref{fig:Y}, when corrections to static propagators are included, and UV divergences arise from the first two topologies in Fig.~\ref{fig:3pnUV}. The divergent terms can be recast in the following form,
\bea
{\cal L}^{\rm IR/UV\, (near)}_{\rm 3PN} &=& -\frac{11}{3} \left(\frac{1}{\epsilon_{\rm IR}}-2 \log{\bar\mu r}\right)G^2m_1^2m_2  \left(\ba_1^2 + 2 \ba_1\cdot \ba_2 \right)\label{UV3pn} \\
&&+\frac{11}{3} \left(\frac{1}{\epsilon_{\rm UV}}-3 \log{\bar\mu r}\right) \left(\frac{G^3m_1^3m_2 }{r^3} \ba_1\cdot\br \right) +( 1 \leftrightarrow 2),
\eea
where we have kept the near zone logarithms, written in terms of $\bar\mu$ given in \eqref{barmu}, which we will use throughout the computations. At 3PN order, the UV pole depends on the dynamical variables and therefore it cannot be simply absorbed into a mass renormalization. At the same time, the IR pole does not have either the structure of a multipole moment which can be associated with computations in the long-distance theory. Both these issues are connected. The resolution relies again on the inclusion of self-energy contributions, which are required to identify the coefficients of the IR/UV poles. The contribution from all of the logarithmically divergent scaleless integrals,  given at 3PN also by the topology in the first diagram of Fig.~\ref{fig:Y3}, yields\footnote{The factors of $r^{2\epsilon}$ account for the correct units of the worldline Lagrangian (after using Newton's constant in $d$ dimensions, see \eqref{eq:gd}). The extra constants in the definition of $\bar\mu$ (see \eqref{barmu}), which the reader will notice cancel between the two terms, are introduced solely  for convenience.}
\beq
{\cal L}^{\rm IR/UV\,(self)}_{\rm 3PN} = -\frac{11}{3}G^2 \left(m_1^3 \ba_1^2 +  m_2^3 \ba^2_2 \right)\left( \frac{(\bar\mu r)^{-2\epsilon_{\rm IR}}}{\epsilon_{\rm IR}}  -\frac{(\bar\mu r)^{-2\epsilon_{\rm UV}}}{\epsilon_{\rm UV}} \right)\,.
\label{eq:selfIR3pn}
\eeq
For the total result we then find
\bea
\label{eq:div3pn}
{\cal L}^{\rm IR/UV\, (near+self)}_{\rm 3PN} &=& -\frac{11}{3}  \left(\frac{1}{\epsilon_{\rm IR}}-2 \log{\bar\mu r}\right) M_0 G^2 \left(m_1 \ba_1 + m_2 \ba_2\right)^2\\
&&+ \frac{11}{3}\left[ \left(\frac{1}{\epsilon_{\rm UV}}-2 \log{\bar\mu r}\right)G^2 m_1^3 \ba_1\cdot \left(\ba_1+\frac{Gm_2\br}{r^3}\right) \right.\nn\\
&&-\left.\frac{G^3 m_1^3m_2\ba_1\cdot\br}{r^3}\log \bar\mu r+ (1\leftrightarrow 2) \right]  \,.\nn
\eea
The reader will now easily notice (as we demonstrate momentarily) that the near zone UV divergence can be removed by a counter-term whose operator vanishes on-shell. (That is the case because it is proportional to the leading equation of motion: $\ba_1+\tfrac{Gm_2\br}{r^3}=0$.) In turn, this implies that physical results are independent of the renormalization-scheme. Moreover, it is also straightforward to see that the (unphysical) IR pole has the form of a dipole-radiation coupling. We will show in the next section it is linked to a UV divergence arising from the computation of radiation-reaction effects in the long-distance EFT. In addition, we will also demonstrate how the logarithmic terms, associated to both IR and UV poles, are removed from physical quantities.

\subsubsection*{4PN order}

The computation of the near zone local-in-time gravitational potential at 4PN order was carried out in \cite{Foffa:2019rdf}. Similarly to the computations in harmonic gauge in \cite{Bernard:2015njp}, which are shown to be equivalent, the divergent terms were reported in \cite{Foffa:2019rdf} as poles in $(d-3)$, without distinguishing their IR or UV nature. However, as we have repeatedly emphasized, the proper renormalization of the effective theory relies on the correct identification of the IR and UV divergences. As we argued, this requires not only isolating the nature of the singularities in the Feynman integrals computed in \cite{Foffa:2019rdf}, but also to incorporate scaleless self-energy diagrams, as in Fig.~\ref{fig:Y3}. After adding all the relevant diagrams including self-energy contributions, which at 4PN entail also the second and third topologies in Fig.~\ref{fig:Y3}, we find:
\allowdisplaybreaks
\bea
\label{4PN_G2pole}
{\cal L}^{{\rm IR/UV\, (near+self)}}_{G^2,\,{\rm 4PN}}&=&- G^2 m_1^2 m_2\left[\frac{2}{15}(\bb_1\cdot\br) (\bb_2\cdot\br)-\frac{19}{15} r^2(\bb_1\cdot \bb_2) 
+\frac{34}{15}(\bv\cdot \ba_1)(\bb_2\cdot\br)\right. \\ &+&\frac{12}{5}(\bv\cdot\ba_2)(\bb_1\cdot\br)+(\ba_2\cdot \bb_1)\left(\frac{12}{5} \bv_1\cdot \br +\frac{74}{15}\bv_2\cdot \br\right)+\left(\frac{19}{15}  \ba_2\cdot \br+\frac{11}{6} \bv^2\right) \ba_1^2\nonumber\\
&+&   \left(\frac{14}{3} \ba_1\cdot \br-\frac{34}{15} \ba_2\cdot \br+\frac{134}{15}\bv_1^2-\frac{16}{5} \bv_1\cdot \bv_2+\frac{79}{15}\bv_2^2\right) (\ba_1\cdot \ba_2)+\frac{11}{2}\bv_1^2\ba_1^2\nonumber\\
&+&\left.\frac{64}{5} (\bv\cdot \ba_1)( \bv\cdot \ba_2)+\frac{22}{3}(\bv_1\cdot \ba_2) (\bv_2\cdot \ba_1)+\frac{11}{3}(\bv_1\cdot \ba_1)^2\right]\left(\frac{1}{\epsilon_{\rm IR}}-2 \log{\bar\mu r}\right)
 \nn \\ &-& 
\frac{11}{3}G^2 m_1^3\left[\left(\frac{1}{\epsilon_{\rm IR}}-2 \log{\bar\mu r}\right)-\left(\frac{1}{\epsilon_{\rm UV}}-2 \log{\bar\mu r}\right)\right]\left((\bv_1\cdot \ba_1)^2+\frac32\bv_1^2\ba_1^2\right)+(1\leftrightarrow 2)\,,\nonumber\\
\label{4PN_G3pole}
{\cal L}^{{\rm IR/UV\, (near+self)}}_{G^3,\,{\rm 4PN}}  &=&\frac{G^3 m_1^3 m_2}{r^3} \left[r^2 \left(\frac{55}{3} \ba_1^2-\frac{53}6 \ba_1\cdot \ba_2 \right)-\frac{11}6( \ba_1\cdot\br)( \ba_2\cdot\br)-11 (\bv\cdot \br) (\bv\cdot \ba_1)\right. \\
&+&\left. \frac{11}{3}(\bv_1\cdot \br)(\bv_2\cdot \ba_1)  + (\ba_1\cdot\br) \left(\frac{22}{3} \bv^2 + \frac{11}6 \bv_1^2-\frac{11}2\frac{(\bv_2\cdot \br)^2}{r^2}\right)\right]\left(\frac{1}{\epsilon_{\rm UV}}-3 \log{\bar\mu r}\right)\nn \\
&+&\ \frac{G^3 m_1^3 m_2}{r^3}
\left[\left(\frac{124}{15}\ba_1^2+\frac{10}{3}\ba_1\cdot \ba_2\right)r^2+\frac{16}{5}{(\ba_1\cdot \br)}^2+\frac{14}{3}(\ba_1\cdot \br) ( \ba_2\cdot \br)\right.\nn\\
&+&\left.\ba_1\cdot \br \left(-\frac{14}{3}\bv^2+14 \left(\frac{\bv\cdot \br}{r}\right)^2\right)-\frac{4}{3}(\bv\cdot \br)(\bv\cdot \ba_1)\right]\left(\frac{1}{\epsilon_{\rm IR}}-3 \log{\bar\mu r}\right)\nonumber\\
&+&\ \frac{G^3 m_1^2 m_2^2}{r^3}\left[\left(\frac{10}{3}\ba_1^2+\frac{124}{15}\ba_1\cdot \ba_2\right)r^2+\frac{10}{3}{(\ba_1\cdot \br)}^2+\frac{68}{15}(\ba_1\cdot \br)(\ba_2\cdot \br)\right.\nn \\
&-& \left. \frac{14}{3} \ba_1\cdot \br \left(\bv^2-3\left(\frac{\bv\cdot \br}{r}\right)^2\right)\right]\left(\frac{1}{\epsilon_{\rm IR}}-3 \log{\bar\mu r}\right) + (1\leftrightarrow 2)\,,\nonumber\\
\label{4PN_G4pole}
{\cal L}^{{\rm IR/UV\, (near+self)}}_{G^4,\,{\rm 4PN}}&=&\left(\frac{1}{\epsilon_{\rm UV}}-4 \log{\bar\mu r}\right) \frac{G^4 m_1 m_2}{3r^4} \big(11m^2_1m_2 - 23 m_1^3\big) \ba_1\cdot \br\\
&+&\left(\frac{1}{\epsilon_{\rm IR}}-4 \log{\bar\mu r}\right) \frac{4G^4 m_1^3 m_2^2}{r^4} \bv^2 + (1\leftrightarrow 2)\nn\,,
\eea
where, for future reference, we have organized the results in powers of~$G$.\footnote{Notice the pole structure reported in \cite{Foffa:2012rn,Foffa:2019rdf} at ${\cal O}(G^2)$ is somewhat different than the one given here. Moreover, all of the logarithmic terms are also different. This, of course, has no physical consequences and it is entirely due to the use of integration by parts and double-zero tricks.} The divergences at ${\cal O}(G^5)$ cancel out in the final result,~see~\cite{Foffa:2016rgu, Damour:2017ced}.
\vskip 4pt 
In the next section we discuss the removal the UV poles using the point-particle effective action in \eqref{eq:spp}. This can be achieved, as expected, without knowledge from the radiation zone. We will show how to handle the IR singularities in the subsequent section, where we also discuss the contributions due to radiation modes, which are UV divergent. The main difference at 4PN order, with respect to the 2PN and 3PN case, is the emergence (after the cancellation of these spurious IR/UV poles) of local- as well as nonlocal-in-time physical effects in the dynamics.

\subsection{UV Counter-terms} \label{sec:counter}

\subsubsection*{Operator basis}

\begin{figure}[!t]
\centerline{\scalebox{0.8}{\includegraphics{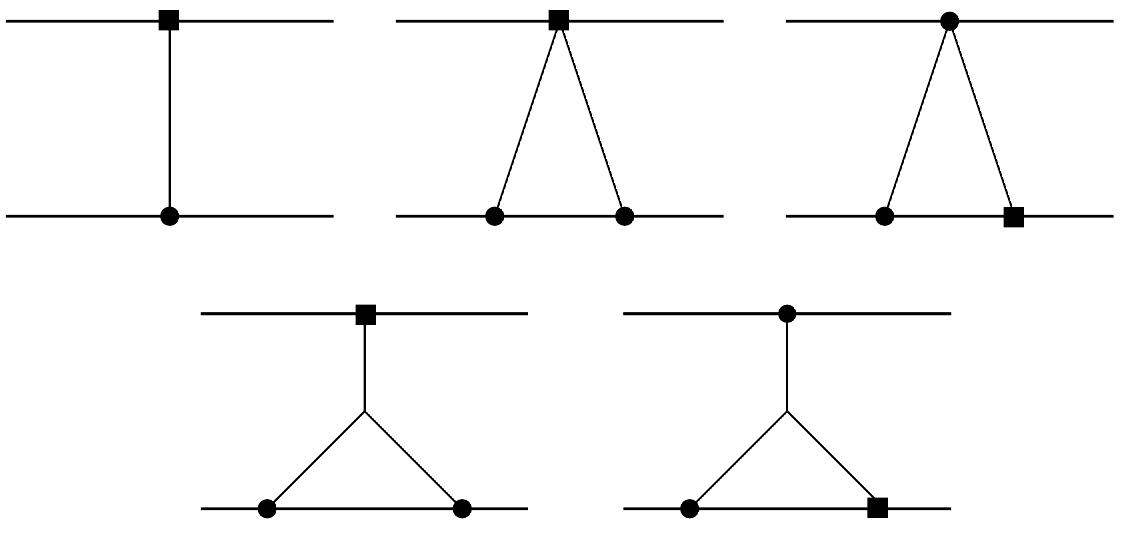}}}
\caption[1]{Topologies for the two counter-terms required to remove divergences to 4PN. The black square represents an insertion of either ${\cal O}_{a\dot v}$ or ${\cal O}_V$ (see text). Mirror images are also needed.} \label{fig:ct}
\end{figure}

Our task now is to remove the UV divergence from the near zone by adding the appropriate counter-term. We will use the following basis of operators in \eqref{eq:spp}:
\beq
\Big\{ {\cal O}_{a\dot v}, {\cal O}_R, {\cal O}_V\Big\}\,.\label{opbasis}
\eeq  
The contributions to the effective action induced by these higher-derivative terms is obtained by including new vertices in the Feynman diagrams. Notice that the first term introduces also corrections which are purely kinematic, namely they do not depend on the variables associated with the other body. (Although, needless to say, the acceleration of each body depends on the existence of a gravitational pull induced by the companion.) The topologies needed to compute the contributions from counter-terms are displayed in Fig.~\ref{fig:ct}. The coefficient for these operators are chosen such that, at each PN order,
\beq
{\cal L}^{\rm UV\, (near+self)}_{n{\rm PN}} + {\cL}^{\rm c.t.\, (near)}_{n{\rm PN} } \to {\rm UV \, finite}\,.
\eeq
As it was shown in \cite{Foffa:2016rgu,Damour:2017ced}, the divergences at ${\cal O}(G^5)$ cancel out at 4PN. It is easy to see that  ${\cal O}_{a\dot v}$ does not generate a contribution at such order in $G$. Moreover, we can also show that neither does the ${\cal O}_V$ operator. On the other hand, ${\cal O}_R$ contributes at ${\cal O}(G^5)$ in harmonic gauge. This means that it must have a finite $c_R$ coefficient in the $d\to 3$ limit, to avoid introducing unaccounted UV poles. Such finite piece may be removed by a field-redefinition, and therefore it plays no role in the renormalization of the theory. Hence, to deal with the UV divergences we must simply fix two parameters: $(c_{a\dot v}, c_V)$.\footnote{This is in contrast to what was found in \cite{nrgr} working in background-field gauge, where $c_R$ and $c_V$ are needed to remove the poles in the one-point function. We will postpone for future work the issue of on-shell vs. off-shell diffeomorphism invariance in the two-body problem, which arises due to the introduction of the $c_{a\dot v}$ coefficient in (standard) harmonic gauge.} Remarkably, ${\cal O}_V$ starts to contribute at 4PN order, leaving only one of them (other than the mass) to renormalize the theory to 3PN order. Because of Lorentz invariance, which relates different PN orders, we shall see how determining $c_{a\dot v}$ at 3PN readily resolves higher order divergences at 4PN, up to a small mismatch which is fixed by the $c_V$ coefficient.

\subsubsection*{Determination of $c_{a\dot v}$}

 The contribution from the ${\cal O}_{a\dot v}$ operator in $d$ dimensions is obtained after expanding in the PN regime. For instance, for particle 1,
\bea
\label{Oavop}
 c_{a\dot v} \int {\cal O}_{a\dot{v}}\ {\rm d}{\tau}&\to&c^{(1)}_{a\dot v} \int {\rm d}t\left[\ba_1^2\left(1+\frac 32\bv_1^2-(1+c_d)\phi\right) \right.\\
 &+&  (\ba_1\cdot\nabla\phi)\left(1+\frac{(1+c_d)}{2}\bv_1^2-\phi\right) \nn \\
&+& (\bv_1\cdot\ba_1)\left((\bv_1 \cdot \ba_1)+(1-c_d)\left(\dot\phi+\bv_1\cdot\nabla\phi\right)\right)\nn \\ 
&+& \left. (\dot {\bA}\cdot\ba_1)+(\bv_1\cdot\nabla)({\bA}\cdot\ba_1)-v_1^ia_1^j\nabla_j\bA_i+\dots\right]\nn\,,
\eea
with $c_d\equiv\frac{2(d-1)}{(d-2)}$ and $\nabla_i$ represents the covariant derivative. The ellipses stand for terms involving interactions which are not relevant at 4PN. We have used the decomposition into scalar, $\phi$, vector, $\bA$, and tensor, $\sigma$, modes, see e.g. \cite{Foffa:2016rgu}. The correction to the Lagrangian first enters at 3PN order and it takes the form, for particle 1,
\be
{\cL}^{c^{(1)}_{a\dot v},\,({\rm near})}_{\rm 3PN}=c^{(1)}_{a\dot v}\, {\ba_1}.\left[{\ba_1}+\frac{Gm_2{\br}}{r^3}\left(1-\epsilon_{\rm UV}\left(\log{\bar\mu r}-\frac{3}{2}\right)\right)\right] \label{Ladv}\,,
\ee
and similarly for particle 2. The logarithmic term plus a constant piece, entering at ${\cal O}(\epsilon_{\rm UV})$, are important to ensure the vanishing on-shell of the ${\cal O}_{a\dot v}$ operator in $d$ dimensions. Notice only the kinematic part contributes at leading order. It is straightforward to show that the UV divergence in \eqref{eq:div3pn} can be easily removed by choosing:
\be
c_{a\dot v, {\rm c.t.}}^{(a)}= -\frac{1}{\epsilon_{\rm UV}} \frac{11}{3} G^2 m_a^3\,.\label{cadv}
\ee
This is in essence equivalent to the worldline shift introduced in \cite{Blanchet:2003gy}. However, the virtue of working at the level of the action is that, once their coefficients are fixed, the counter-terms are determined to all orders by symmetries, in our case locality and Lorentz invariance. 
\subsubsection*{Determination of $c_V$}
Incorporating the value obtained in \eqref{cadv}, the contribution from $c_{a\dot v, {\rm c.t.}}$ to the effective action at 4PN then reads, see \eqref{Oavop},
 \bea
 \label{Ladv2}
{\cL}^{c_{a\dot v,{\rm c.t.}}, \rm (near)}_{\rm 4PN}  &=& -\frac{1}{\epsilon_{\rm UV}} \frac{11}{3} G^2 m_1^3\left(\frac 32 \ba_1^2 \bv_1^2+\left(\ba_1\cdot \bv_1\right)^2\right)\\
&-&  \frac{G^3 m_1^3 m_2}{r^3} \Bigg[ \frac{11}{3}\left(\frac{1}{\epsilon_{\rm UV}}-(\log \bar\mu r-1)\right) \left(r^2 \left(5 \ba_1^2-\frac72 \ba_1\cdot \ba_2 \right)-\frac12( \ba_1\cdot\br)( \ba_2\cdot\br)\right.\nn \\ &+&\left. (\bv_1\cdot \br)(\bv_2\cdot \ba_1)+ (\ba_1\cdot\br) \left(2 \bv^2 + \frac12 \bv_1^2-\frac32\frac{(\bv_2\cdot \br)^2}{r^2}\right) -3 (\bv\cdot \br) (\bv\cdot \ba_1)\right)\nn\\
&+& r^2 \left(\frac{15}{4} \ba_1\cdot \ba_2-\frac92 \ba_1^2\right) -\frac14( \ba_1\cdot \br) ( \ba_1\cdot \br)+ (\ba_1\cdot \br) \left(\frac14 \bv_1^2 -\frac54\left(\frac{\bv_2\cdot \br}{r}\right)^2\right)\nn \\ &+& \frac12 (\bv\cdot \br)(\bv_1\cdot \ba_1)+\frac12 (\bv_2\cdot \br)(\bv_2\cdot \ba_1)\Bigg]\nn \\ &+& \frac{11}{3}\Big(\frac{1}{\epsilon_{\rm UV}}-2\left(\log{\bar\mu r}-1\right)\Big)\frac{G^4 m_1^3 m_2(m_1-m_2)}{r^4}(\ba_1\cdot \br ) + \left(1\leftrightarrow 2\right)\,.\nn
\eea
The above expression already knocks off many of the UV divergences in \eqref{4PN_G2pole} - \eqref{4PN_G4pole}. It is then straightforward to show that the remaining UV poles can be absorbed into the ${\cal O}_V$ operator. The contribution to the effective action takes the form 
\bea
{\cL}^{c_V,\,({\rm near})}_{\rm 4PN}&=&4 c^{(1)}_{V}\left(- \ba_1\cdot \ba_2  \frac{G m_2}{r}\left(1-\epsilon_{\rm UV}\log{\bar\mu r}\right)\right.\\ &+& \left.(\ba_1\cdot \br) \frac{G^2 m_1 m_2}{r^4}\left[1-2\epsilon_{\rm UV}\left(\log{\bar\mu r}-\frac34\right)\right]\right)+\left (1\leftrightarrow 2\right)\,,\nn
\eea
such that the choice of counter-term
\beq
\label{cv}
c^{(a)}_{V, {\rm c.t.}} = \frac{1}{\epsilon_{\rm UV}} G^2 m_a^3\,,
\eeq
removes the remaining UV poles. In summary, the (bare) point-particle effective action in \eqref{eq:spp} becomes\footnote{Notice that the, rather simple, expression in \eqref{eq:spp2} predicts much of the structure of UV poles that will appear also at higher PN orders. In~contrast, in the Fokker-action treatment the removal of the UV divergences is performed, independently at each PN order, through an {\it ad hoc} worldline shift \cite{Bernard:2015njp}.}
\bea
\label{eq:spp2}
S_{\rm pp}[x^\alpha_a(\tau_a)] &=& \sum_a \int d \tau_a \left[-m_a +\left(c^{(a)}_{a\dot v,\,{\rm ren}}(\mu) - \frac{11}{3}\frac{G^2 m_a^2}{\epsilon_{\rm UV}} \right) g_{\mu\nu} a_a^\mu \dot v_a^\nu\right. \\ &+& \left. \left(c^{(a)}_{V,\,{\rm ren}}(\mu) + \frac{G^2 m_a^2}{\epsilon_{\rm UV}}\right) R_{\mu\nu}v_a^\mu v_a^\nu \right]\,.\nn
\eea
This completes the UV renormalization of the theory to 4PN order.\vskip 4pt While we have chosen a MS scheme in \eqref{cadv} and \eqref{cv} to remove the poles, there is always a degree of ambiguity in the choice of counter-terms. (For instance, we could choose the $\overline{\rm MS}$ scheme and remove the extra constants we kept in the definition of $\bar\mu$.) This is often resolved by a matching computation, as we alluded before. However, as we discussed, all of the operators in \eqref{opbasis} can be removed by field-redefinitions.  This means that shifts in the counter-terms or renormalized parameters have no physical effect. We will return in sec.~\ref{sec:renormtot} to address this remaining freedom in the effective theory. As we shall see, the renormalized coefficients can be used to remove unphysical UV logarithms (as well as factors of $\mu$) from the final expressions. 

\section{Cancellation of near/far zone IR/UV divergences}\label{sec:cancel}

After the near zone UV divergences are removed by counter-terms, the remaining task is to deal with the IR singularities. As we discussed, this is done by implementing the zero-bin subtraction \cite{Porto:2017dgs,Porto:2017shd}. In practice, the subtraction of the zero-bin transforms IR into UV poles (see \eqref{eqzerob} below). The main reason is due to the type of integrals that enter at this order, such that the zero-bin appears as a scaleless contribution \cite{Porto:2017dgs}. Once the IR poles turn into UV divergences, they cancel out against UV poles arising in conservative contributions from the far zone. The divergences appear in the computation of the tail effect in the long-distance effective theory, in which the binary itself is treated as a point-like source endowed with a series of multipole moments. While the subtraction of the zero-bin may seem like a simple formal manipulation in dim. reg., it determines unambiguously all the finite pieces in the gravitational potential without room for ambiguities. Moreover, the procedure is regularization-independent \cite{zb}, and therefore it may be used to remove the ambiguities plaguing other derivations. As we demonstrate, the cancellation between spurious divergences is already at work at 2PN and 3PN orders. Starting at 4PN, tail terms from the radiation region begin to contribute to the conservative dynamics.

\subsection{Zero-bin IR subtraction}

The IR divergences which we encounter when computing in the potential region can be ultimately traced down to the master integral (similar to the so-called `Riesz formula' in $d$ dimensions \cite{Schafer:2018kuf})
 \begin{align}
 \label{footl}
 I[n_1,n_2]&\equiv \int_\bk \frac{1}{[\bk^2]^{n_1}\left[(\bk+\bp)^2\right]^{n_2}} \\ &= \frac{\Gamma[n_1+n_2-d/2]\Gamma[d/2-n_2]\Gamma[d/2-n_1]}{(4\pi)^{d/2}\Gamma[n_1]\Gamma[n_2]\Gamma[d-n_1-n_2]} \left(\bp^2\right)^{d/2-n_1-n_2}\,,\nn
 \end{align}
where $\int_\bk \equiv \int \tfrac{d^d\bk}{(2\pi)^d}$, see appendix~\ref{sec:app0}. The IR divergence is manifest by the fact that the right-hand-side can become singular for $d<3$. As discussed in \cite{Porto:2017dgs}, the implementation of the zero-bin subtraction is straightforward in this case. The singular term comes from the region of integration where the momenta is soft: $\bk \ll \bp$, which must be subtracted away. For instance, the pair $(n_1=3/2,n_2=1/2)$ occurs repeatedly. In that case, expanding the factor of $(\bk+\bp)$, the zero-bin contribution becomes\footnote{In $d$ dimensions, the (scaleless) zero-bin integral(s) includes also the additional factors displayed in e.g. \eqref{eq:selfIR3pn}.} \cite{Porto:2017dgs}
\beq
I_{\rm ZB} \left[n_1,n_2\right] = \int_\bk \frac{1}{[\bk^2]^{n_1}[\bp^2]^{n_2}} \xrightarrow{(n_1= 3/2,n_2 = 1/2)} 
|\bp|^{-1} \int_\bk \frac{1}{\bk^3} = \frac{i}{16\pi |\bp|} \left(\frac{1}{\epsilon_{\rm UV}} - \frac{1}{\epsilon_{\rm IR}}\right)\,. \label{zerobin0}
\eeq
It is now straightforward to implement the subtraction in all of the IR divergent Feynman integrals to 4PN order, {\it including also the self-energy contributions}, which themselves are scaleless-type integrals. At the end of the day, after removing the zero-bin, in practice the IR poles turn into UV divergences according to the replacement, for $n\leq 4$,

\beq
{\cal L}^{\rm UV\, (IR\, near+self-ZB)}_{n{\rm PN}} \equiv {\cal L}^{\rm IR\, (near+self)}_{n{\rm PN}}{\left. \right|}_{\epsilon_{\rm IR} \to \epsilon_{\rm UV}}\,.\label{eqzerob}
\eeq

Let us stress two important points. First of all, while the subtraction of the zero-bin changes the nature of the pole, {\it crucially} it does not introduce any extra finite pieces at this order. This feature, which is only true in dim. reg., it is also due to the type of integrals that contribute to 4PN order. Secondly, the UV poles from the scaleless integrals have been absorbed into counter-terms, as expected. Therefore, there are left-over IR divergences from self-energies in the near zone, turning into UV poles after the zero-bin is removed from the scaleless integrals. In what follows we demonstrate how the left-over UV poles, after the subtraction of the zero-bin, explicitly cancel out against counter-parts arising from conservative contributions in the far zone.\footnote{Note that algebraically, if applied before the UV counter-terms, the zero-bin subtraction would cancel self-energy terms exactly. Of course, at the end of the day, the remaining (artificial) UV poles in the near and far zones would also vanish in the final results. However, the procedure outlined here, working with \eqref{eq:spp2} to  renormalize the near zone effective theory with potential modes before implementing the zero-bin subtraction, clarifies the nature of the spurious poles and demonstrates the mutual cancellation. See sec.~\ref{sec:disc} for more details.}

\subsection{Radiation-reaction UV poles}\label{secfarz}

In the calculation of the tail contribution yielding \eqref{eq:RRnl} at 4PN order, only the physical quadrupole coupling in \eqref{eq:srad} was retained. However, as we emphasized, other type of couplings (e.g. the dipole term) may induce UV divergences in the far zone, albeit without leading to a physical effect. Below we compute all of the UV singularities which appear in the EFT computation of the conservative radiation-reaction tail effects, including those which vanish on-shell. These will be essential to remove the UV poles remaining in the near region after the zero-bin subtraction. For simplicity, all of the results quoted below are given in standard (as opposite to $\pm$) variables, such that the variation of the action follows the usual~steps.

\subsubsection*{2PN}

At 2PN order we encounter a diagram similar to Fig.~\ref{nltail}, but instead of the quadrupole moment we insert the  monopole term, $\int M(\tau) d\tau$ (see \eqref{eq:srad}). Keeping only the conservative part we find:
\bea
\label{eq:uvfar2pn}
\int dt\, \cL^{\rm UV\, (far)}_{\rm 2PN} &=& - i \frac{M_0}{\Mp^2}\int_{\bk,\bq} \int \frac{d\omega}{2\pi} |M(\omega)|^2\big\langle \phi(\omega,\bq)\phi(\omega,\bq-\bk)\phi(0,\bq)\big\rangle_{\rm UV}\\ 
&=& - \frac{G^2M_0}{\epsilon_{\rm UV}} \int \frac{d\omega}{2\pi} \omega^2 |M(\omega)|^2 \left(1 + \epsilon_{\rm UV} \log \left(\frac{4\omega^2\,e^{2\gamma_E}}{\bar\mu^2}\right) \right)\,,\nn
\eea
where, in order to illustrate the cancellation against the near zone logarithms we have written the result in terms of $\bar\mu$ in \eqref{barmu} (but omitted other constants), and already transformed from $\pm$ into standard variables.
While we have taken the leading order part of the background geometry proportional to $M_0$, we have allowed for higher order PN corrections to the monopole coupling, i.e. $M(t)= M_0 + \cdots$, which will play a key role later on.\vskip 4pt 

The reader will immediately notice that the UV pole from the far zone has (minus) the coefficient of the IR counter-part at 2PN in \eqref{eq:div2pn}. Hence, following the zero-bin subtraction, we thus arrive at 
\beq
\int dt \left({\cal L}^{\rm UV\, (IR\, near+self - ZB)}_{\rm 2PN} + \cL^{\rm \, UV\, (far)}_{\rm 2PN}\right) \to - \, G^2 M_0\int \frac{d\omega}{2\pi}  \omega^2 |M(\omega)|^2 \log \left(4\omega^2 e^{2\gamma_E}\right)\,.
\eeq
Needless to say, since $\dot M=0$ on-shell, all these manipulations involving the monopole coupling do not contribute to anything physical. However, we already start to unfold the pattern that will continue to appear at higher PN orders.
\subsubsection*{3PN}

Next, at 3PN order we have corrections from the monopole term as well as contributions from the dipole coupling. The result can be split into two parts. While the effects due to the background geometry remain as the (leading order) scalar exchange, there are now a scalar, $\phi^3$, as well as vector, $A^2\phi$, coupling. The computation is straightforward and, using $\bP = M_0\dot\bX$ at leading order, we find
\bea
\int dt \, \cL^{\rm \, UV\, (far)}_{\phi^3,\, {\rm 3PN}} &=& -\frac{G^2 M^3_0}{3\epsilon_{\rm UV}} \int \frac{d\omega}{2\pi} \omega^4|\bX(\omega)|^2\left( \frac{1}{\epsilon_{\rm UV}} +\log \left(\frac{4\omega^2\,e^{2\gamma_E}}{\bar\mu^2}\right) \right)\,,\\
\int dt\, \cL^{\rm \, UV\, (far)}_{\phi A^2,\, {\rm 3PN}} &=& \frac{4G^2 M^3_0}{\epsilon_{\rm UV}} \int \frac{d\omega}{2\pi} \omega^4|\bX(\omega)|^2\left( \frac{1}{\epsilon_{\rm UV}} +\log \left(\frac{4\omega^2\,e^{2\gamma_E}}{\bar\mu^2}\right) \right)\,,
\eea
where we have written, as before, the associated logarithmic contribution in terms of $\bar\mu$. The two terms combined lead to
\beq
\label{eq:uvfar3pn}
\int dt\, \cL^{\rm UV\,(far)}_{\rm 3PN} = \frac{11}{3} M^3_0 G^2 \int \frac{d\omega}{2\pi} \omega^4|\bX(\omega)|^2 \left( \frac{1}{\epsilon_{\rm UV}} + \log \left(\frac{4\omega^2\,e^{2\gamma_E}}{\bar\mu^2}\right)\right)\,.
\eeq
Once again the reader will identify the coefficient of the far zone UV divergence with (minus) the one in the IR pole from the near region in \eqref{eq:div3pn}, after noticing $\ddot\bX = m_1\ba_1+m_2\ba_2$, at leading order. The cancellation of divergences, as well as the associated $\log \bar\mu$'s, follows, 
\beq
\label{eq:XX}
\int dt \left({\cal L}^{\rm UV\, (IR\, near+self - ZB)}_{\rm 3PN} + \cL^{\rm \, UV\, (far)}_{\rm 3PN} \right) \to  \frac{11}{3} M^3_0 G^2 \int \frac{d\omega}{2\pi} \omega^4|\bX(\omega)|^2 \log \left(4\omega^2 e^{2\gamma_E}\right)\,,
\eeq
as anticipated. The reader will notice, at 3PN order, the remaining logarithm multiplies a term which is a double-zero on-shell, and therefore it can be ignored. This will not be the case at 4PN order.\vskip 4pt

The previous examples illustrate how, once the poles are identified and the zero-bin subtraction implemented, the IR/UV spurious divergences, and associated $\log \mu$'s, cancel each other out between near and far zone contributions. Of course, there are no left-over finite terms to 3PN order from this procedure, since these all vanish on-shell. We find the first non-trivial (finite) contribution to the conservative sector from the far zone at 4PN order. The cancellation of divergences, in any case, proceeds in a similar fashion. 

\subsubsection*{4PN}
The contribution from the quadrupole coupling, $I^{ij} E_{ij}$, was computed before, see \eqref{eq:RRnl}. After plugging the form of $I_0^{ij}$ at leading order, we find:
\beq
\label{QQtail}
S^{\rm tail\, (far)}_{{\rm 4PN},\, M(I^{ij}_0)^2} = S_{\rm 4PN}^{\rm tail\, (loc)}+S_{\rm 4PN}^{\rm tail\, (nonloc)},\,
\eeq
where we have isolate the local-in-time,\footnote{The terms in the third line originate from ${\cal O}(\epsilon_{\rm UV})$ corrections to the quadrupole in $d$ dimensions hitting the UV pole. These terms play a key role to arrive to the correct physical expression in the $d\to 3$ limit.}
\bea
S_{\rm 4PN}^{{\rm tail}\, (\rm loc)} &=& -\frac{G^2 M_0^3\nu^2}5 \int dt\, \Bigg\{\left(\frac1{\epsilon_{\rm UV} }-2 \log{\bar\mu r}-\frac{41}{30}\right)\Big(2\, r^2 \bb^2+\frac{2}{3}\left(\br\cdot \bb\right)^2 +18\, \ba^2\bv^2
 \nn \\
&+&  6 (\bv\cdot \ba)^2 -\, 8\left(\br\cdot \bb\right)(\bv\cdot \ba)+12(\ba\cdot \bb)(\br \cdot \bv)+12(\bv\cdot \bb)(\br \cdot \ba)\Big)\nn\\
&+&\left.\frac{4}{9}\left(\br\cdot \bb\right)^2+4 (\bv\cdot \ba)^2 +\, \frac{8}{3}\left(\br\cdot \bb\right)(\bv\cdot \ba)\right.\Bigg\}  \,,
\eea
with $\nu \equiv \frac{m_1m_2}{M_0^2}$ is the symmetric mass ratio, and nonlocal-in-time,
\beq
\label{G2nonloc}
S_{\rm 4PN}^{{\rm tail}\, (\rm nonloc)} \equiv \int dt \, \cL_{\rm 4PN}^{G^2 \log v} =  -\frac{G^2 M_0}{5} \int_{-\infty}^\infty \frac{d\omega}{2 \pi} \,  \omega^6 \, \left|I^{ij}(\omega)\right|^2 \log \left(4\omega^2 r^2{\rm e}^{2 \gamma_E}\right)\,,
\eeq
contributions to the effective action. The nonlocal-in-time part has been obtained earlier in \cite{Blanchet:1987wq}, and emphasized more recently in \cite{Damour:2014jta}. In order to perform the above splitting, we have separated the $\log \mu$ from \eqref{eq:RRnl} (which we re-wrote in terms of $\bar\mu$) and $\log \omega$, and introduced (by hand) the factors of $\log r$.\footnote{For the sake of simplicity, in writing \eqref{G2nonloc} we are using an abuse of notation where the `mixed term', $\int \tfrac{d\omega}{2 \pi} \, \omega^6 \, \left|I^{ij}(\omega)\right|^2 \log r$, replaces the correct expression $\int \, dt\,  I^{ij(3)}(t) I^{ij(3)}(t) \, \log r(t)$.  The superscript ${(3)}$ indicates three time derivatives.} 
This is simply for convenience, so that in this fashion the factor of $\log\bar\mu r$ cancels out against a logarithmic contribution from the near zone.\footnote{Notice that the cancellation leaves behind finite terms, associated with regularization-dependent constants, i.e. $\left(\bar\mu/\mu\right)^2= 4\pi e^{\gamma_E}$. This mismatch is behind the factor of $\log(16\,e^{2\gamma_E})$, in addition to the $\log x$ (with $x\equiv (GM\omega)^{2/3}$), in the expression for the binding energy in a circular orbit at 4PN, see \cite{Damour:2014jta}.}\vskip 4pt

Notice that, without using equations of motion, the UV pole enters at ${\cal O}(G^2)$. Hence, it is clearly not sufficient to cancel against the IR poles from the near region in \eqref{4PN_G2pole} - \eqref{4PN_G4pole}, after the zero-bin subtraction. Moreover, the coefficients of the $G^2$ divergences do not match either. As we demonstrate in what follows, the reason for the mismatch is due to contributions from the long-distance EFT which we have ignored thus far, since they vanish on-shell, see \eqref{eq:con}. However, the additional terms are needed to remove all of the IR/UV spurious divergences. The extra terms in the effective action that result from the multipole expansion in the far zone, but do not contribute to physical quantities, can be written as follows
\bea
\label{eq:multeqm}
\Delta S_{\rm rad}&=& \frac{1}{\Mp} \int dt\left[\frac{1}{2} \left(T^{ij}-\frac{I^{ij(2)}}{2}\right)\bar{h}_{ij}+\frac{1}{2}\left(\left(M^{ij}+M^{ji}\right)-I^{ij(1)}\right)\bar{h}_{0i,j} + \cdots\right]\,. \eea
We use the superscript ${(n)}$ to indicate $n$ time derivatives. The moments of the pseudo stress-energy tensor that appear in the above expression (other than the usual suspects) are given by
\beq
T^{ij}(t) \equiv \int d^3\bx\, {\cal T}^{ij}(t,\bx)\,,\,\, M^{i(L-1)}(t) \equiv \int d^3\bx\, {\cal T}^{0i}(t,\bx) \bx^{L-1}\,.
\eeq
Notice that, as expected, the coefficients in \eqref{eq:multeqm} vanish once conservation laws are enforced, e.g. \eqref{eq:con}. Yet, each of these terms induces extra (divergent) contributions to the tail effect in radiation-reaction. More explicitly, the expression in \eqref{eq:multeqm} gives rise to terms which can be written in a generic form as follows:
\bea\label{eq:tailsonshell}
 \cL^{\rm tail\,  (far)}_{{\rm 4PN},\, \Delta S_{\rm rad} }  =  G^2\int_{-\infty}^\infty \frac{d\omega}{2 \pi} \,  \omega^6 \, \sum_i{\cal M}_i(\omega) \bigg[\frac{1}{ \epsilon_{\rm UV}}  -  2\log{\bar\mu r}+\log{\left(4\omega^2 r^2 {\rm e}^{2\gamma_E} \right)} + \cdots \bigg]\,,
\eea
where the ellipses include finite terms, similarly to the celebrated $41/30$ in \eqref{eq:RRnl}, and the introduction of $\log \bar \mu r$,  as in \eqref{QQtail}, is for convenience. Unlike the contribution from \eqref{QQtail}, each one of the ${\cal M}_i(\omega)$ vanishes on-shell in $d$ dimensions.\vskip 4pt The computation of additional tail terms is lengthy, but straightforward, see Appendix~\ref{sec:appA}. After gathering all the pieces, the resulting UV poles and logarithmic terms take the form: 
\bea
\label{eq:divZB4pnG2} {\cal L}^{\rm UV\, (far)}_{G^2,\,4{\rm PN}} &=& 
G^2 m_1^2 m_2\left[\frac{2}{15}(\bb_1\cdot\br) (\bb_2\cdot\br)-\frac{19}{15} r^2(\bb_1\cdot \bb_2) +\frac{34}{15}(\bv\cdot \ba_1)(\bb_2\cdot\br)\right. \\ &+&\frac{12}{5}(\bv\cdot\ba_2)(\bb_1\cdot\br)+(\ba_2\cdot \bb_1)\left(\frac{12}{5} \bv_1\cdot \br +\frac{74}{15}\bv_2\cdot \br\right)+\left(\frac{19}{15}  \ba_2\cdot \br+\frac{11}{6} \bv^2\right) \ba_1^2\nonumber\\&+&   \left(\frac{14}{3} \ba_1\cdot \br-\frac{34}{15} \ba_2\cdot \br+\frac{134}{15}\bv_1^2-\frac{16}{5} \bv_1\cdot \bv_2+\frac{79}{15}\bv_2^2\right) (\ba_1\cdot \ba_2)\nonumber\\&+&\left.\frac{64}{5} (\bv\cdot \ba_1)( \bv\cdot \ba_2)+\frac{22}{3}(\bv_1\cdot \ba_2) (\bv_2\cdot \ba_1)+\frac{11}{3}(\bv_1\cdot \ba_1)^2+\frac{11}{2}\bv_1^2\ba_1^2\right]\left(\frac{1}{\epsilon_{\rm UV}}-2 \log{\bar\mu r}\right) \nn \\ &+& \frac{11}{3} G^2 m_1^3\left((\bv_1\cdot \ba_1)^2+\frac32\bv_1^2\ba_1^2\right)\left(\frac{1}{\epsilon_{\rm UV}}-2 \log{\bar\mu r}\right)  + (1\leftrightarrow 2) \nn \\
\label{eq:divZB4pnG3} {\cal L}^{\rm UV\, (far)}_{G^3,\,4{\rm PN}} &=& -\frac{G^3 m_1^3 m_2}{r^3}\left[\left(\frac{124}{15}\ba_1^2+\frac{10}{3}\ba_1\cdot \ba_2\right)r^2+\frac{16}{5}{(\ba_1\cdot \br)}^2+\frac{14}{3}(\ba_1\cdot \br) ( \ba_2\cdot \br)\right.\\&+&\left.\ba_1\cdot \br \left(-\frac{14}{3}\bv^2+14 \left(\frac{\bv\cdot \br}{\br}\right)^2\right)-\frac{4}{3}(\bv\cdot \br)(\bv\cdot \ba_1)\right]\left(\frac{1}{\epsilon_{\rm UV}}-3 \log{\bar\mu r}\right) \nonumber\\&-&\ \frac{G^3 m_1^2 m_2^2}{r^3}\left[\left(\frac{10}{3}\ba_1^2+\frac{124}{15}\ba_1\cdot \ba_2\right)r^2+\frac{10}{3}{(\ba_1\cdot \br)}^2+\frac{68}{15}(\ba_1\cdot \br)(\ba_2\cdot \br)\right.\nn \\&-& \left. \frac{14}{3} \ba_1\cdot \br \left(\bv^2-3\left(\frac{\bv\cdot \br}{r}\right)^2\right)\right]\left(\frac{1}{\epsilon_{\rm UV}}-3 \log{\bar\mu r}\right)   + (1\leftrightarrow 2) \nn\\
\label{eq:divZB4pnG4} {\cal L}^{\rm UV\, (far)}_{G^4,\,4{\rm PN}} &=& -\left(\frac{1}{\epsilon_{\rm UV}}-4 \log{\bar\mu r}\right)  \frac{4G^4 m_1^3 m_2^2}{r^4} \bv^2 + (1\leftrightarrow 2) \,.
\eea
where we performed the manipulations to introduce the factors of $\log r$ described after \eqref{G2nonloc} and, for notational simplicity, we did not include the left-over $\log \omega r$'s (see below). The cancellation between the near and far zone divergences as well as $\log \bar\mu r$'s is now evident, \beq
\label{420}
\int dt \left({\cal L}^{\rm UV\, (IR\, near+self-ZB)}_{4{\rm PN}} +{\cL}^{\rm UV\, (far)}_{\rm 4PN }\right) \to \int \frac{d\omega}{2\pi} \left(\cdots \right)\times \log\left( 4\omega^2 r^2e^{2\gamma_E}\right)\,.
\eeq
similarly to the 3PN case in \eqref{eq:XX}. The ellipses representing a series of finite terms at each order in $G$. The result thus includes long-distance logarithms, of which only the one in \eqref{G2nonloc} enters in the conservative dynamics.\vskip 4pt

It is key to notice that, even though they do not play a role for the cancellation of intermediate divergences, there are other extra finite pieces resulting from the terms displayed in \eqref{eq:tailsonshell}, which are required to ensure the cancellation of unphysical (finite) contributions. These terms arise, as in \eqref{QQtail}, from the UV poles hitting the ${\cal O}(\epsilon_{\rm UV})$ corrections in the (vanishing) $d$-dimensional multipole moments (${\cal M}_i(\omega)$) entering in \eqref{eq:tailsonshell}, after performing the matching to the near zone. (See appendix \ref{sec:appA}.) These terms, which can be written in compact form as:
\beq
\label{eq:multeqmfin}
{\cL}^{\rm tail\, (far)}_{\rm 4PN,\, \Delta S_{\rm rad} (finite)}=-4\left[\frac{G^4 m_1^3 m_2}{r^4}\left(5 \bv^2-4\frac{( \bv\cdot\br)^2}{r^2}\right)+\frac{G^5 m_1^3 m_2^2 M_0}{r^5}\right]+\left(1\leftrightarrow 2 \right)+\cdots\,,
\eeq
must be kept in the renormalized Lagrangian (in addition to the extra ones in the third line of \eqref{QQtail}) to arrive to correct (and unambiguous) physical expressions.\footnote{In the Fokker-action approach, using a (short-distance) worldline redefinition to remove the (long-distance) IR poles \cite{Bernard:2015njp}, the `extra terms' are due to ${\cal O}(\epsilon)$ corrections to the equations of motion in $d$ dimensions. Our approach, on the other hand, illustrates the true origin of these terms.}  The ellipses in the above equation account for other (finite) contributions which are zero on lower order equations of motion, and are therefore irrelevant for all physical purposes. 

\section{Effective theory to 4PN order}\label{sec:renormtot}

The renormalization procedure described in this paper, schematically
\beq
\Big( {\cL}^{\rm UV\, (near+self)}_{n{\rm PN}}+{\cL}^{\rm c.t.\, (near)}_{n{\rm PN}}\Big)+\Big({\cal L}^{\rm UV\,(IR\, near+self-ZB)}_{n{\rm PN}} + {\cL}^{\rm UV\, (far)}_{n{\rm PN}}\Big)  \to {\rm finite}\,,\label{eq:cancel}
\eeq
explains how the intermediate IR and UV divergences are removed, or cancel out, from the renormalized effective action at a given $n$PN order. At the end of the day, including all the finite pieces, we are left with a series of local- and nonlocal-in-time contributions to the Lagrangian, which we display momentarily. We have also a series of logarithmic terms, of which only the (long-distance) one shown in \eqref{eq:RRnl} contributes to physical quantities.

\subsection{Long-distance logarithms}

In addition to the poles, we have also demonstrated how the factors of $\log \mu$, associated to IR divergences in the potential region, cancel out against conservative UV logarithms from the tail integrals in the far zone. This is not surprising, and it is entirely due to the general rule,
\beq\label{polelogs}
G_d^n/\epsilon \equiv \left(\mu^{-\epsilon}G\right)^n/\epsilon \to G^n \left( 1/\epsilon - n \, \log \mu\right) \,,  
\eeq
which links the poles to the factors of $\log \mu$ in dim. reg., at a given $n^{\rm \small th}$ order in $G$.\vskip 4pt As we argued, while the $\log \mu$'s disappear, long-distance logarithms of the form~$\log \omega r$ remain in the Lagrangian. For instance, the dipole term at 3PN, see \eqref{eq:XX}. This correction in particular is proportional to a term which vanishes on-shell, and therefore it does not contribute to physical quantities. However, for the case of the quadrupole term in \eqref{eq:RRnl} at 4PN, the key difference is that the associated $\log \omega r$ is no longer proportional to a term which vanishes on-shell, due to a conservation law. For all of the other corrections induced by \eqref{eq:multeqm}, which vanish upon using moment relations, we can show that there is no observable contribution provided all the relevant terms are included (see \eqref{eq:multeqmfin}).

\subsection{Short-distance logarithms}

There are also factors of $\log \bar\mu r$ associated to UV divergences in the potential region. They are as well as due to \eqref{polelogs}, and the expansion of the $d$-dimensional Green's (as well as $\Gamma$) functions around $d=3$ in dim. reg. These logarithms remain after removing the UV poles through counter-terms, e.g \eqref{cadv}, while the finite pieces are renormalization-scheme dependent. The $\mu$-dependence, on the other hand, is absorbed into renormalized coefficients, $c_{\alpha,\, {\rm ren}}(\mu)$, such that we have a renormalization group equation, 
\beq
\mu \frac{d}{d\mu} c_{\alpha,\, {\rm ren}}(\mu) = \beta_\alpha \, G^n m^{n+1}\,,
\eeq
following from the condition $\mu \frac{d}{d\mu}\cL=0$, and the general structure \beq
\cL = \cdots + \left( c^{(a)}_{\alpha,\, {\rm ren}}(\mu) - \beta_\alpha G^n m_a^{n+1} \log\mu r +\cdots \right)\times f_\alpha(\bx_a,\bv_a,\bb_a,\cdots)+\cdots\,,
\eeq
of the renormalized Lagrangian. In general, these type of logarithms contribute to physical quantities, and the renormalization group equation allows us to resum many of such contributions,~e.g.~\cite{andirad,Galley:2015kus}. However, for our case at hand, the fact that the $c^{(a)}_{\alpha,\, {\rm ren}}(\mu)$ coefficients can be removed by field-redefinitions implies that, likewise, the logarithmic running does not contribute to physical quantities. (In other words, $f_\alpha(\bx_a,\bv_a,\bb_a,\cdots)$ vanishes on-shell to 4PN.)\vskip 4pt

To illustrate the situation, let us consider once again the 3PN case. After implementing the MS scheme, we have\footnote{We could have equally used the $\overline{\rm MS}$ scheme, and subtract the UV poles plus all of the constants going into the definition of $\bar\mu$ in \eqref{barmu}. The difference is an inconsequential shift in the renormalized parameters.} 
\beq
\label{eq:logtot}
\cL^{\rm UV\, (near+self)}_{\rm 3PN}  + {\cL}^{c_{a\dot v},\,({\rm near})}_{\rm 3PN}= \left(c^{(1)}_{a\dot v,\,{\rm ren}}(\mu) - \frac{22}{3} G^2 m_1^3 \log \bar\mu r  +\cdots \right) \ba_1\cdot \left(\ba_1+\frac{G m_2 \br }{r^3}  \right)  + ( 1\leftrightarrow 2)\,,
\eeq
after adding \eqref{eq:div3pn} and \eqref{Ladv}, and splitting the bare coefficient $c^{(1)}_{a\dot v}$ into a counter-term, given in \eqref{cadv}, plus $c^{(1)}_{a\dot v,\,{\rm ren}}(\mu)$, the renormalized piece.\footnote{Since the counter-term lives in the (one-dimensional) worldline, there are no factors of $\mu$ associated with $c_{a\dot v}$ in $d$ bulk dimensions. Notice, however, the logarithmic term in \eqref{Ladv}, due to the $d$-dimensional Green's function, are essential to obtain the form in \eqref{eq:logtot}. The same for the logarithmic contributions in \eqref{Ladv2}~at~4PN.} 
It is now straightforward to show that the choice\footnote{Formally speaking, the variation of the action must be performed prior to choosing the value of $\mu$. However, only the variation of $f_\alpha(\cdots)$ contributes. Therefore, setting $\mu$ in the action is de facto innocuous.} 
\beq \mu = (r \sqrt{4\pi})^{-1} e^{-\gamma_E/2}\,,\eeq removes the logarithmic contribution, as well as some associated constants. The logarithmic contributions are then encoded in the running of the renormalized coefficient. Yet, the entire term multiplies a factor that is proportional to the leading order equation of motion, and therefore it can be removed from physical quantities. In our language, the $c_{a\dot v}$ coefficient can be set to zero by a field redefinition, thus erasing all information about factors of $\log\bar\mu r$.

\subsection{Renormalized Lagrangian}

Below we quote the final expression for the renormalized Lagrangian to 4PN order, using the intermediate results reported in \cite{Foffa:2011np,Foffa:2012rn,Foffa:2016rgu,Foffa:2019rdf,Galley:2015kus}, together with the procedure described in~\eqref{eq:cancel}~\cite{Porto:2017dgs,Porto:2017shd}. The reader will find no trace of $\log \mu$ (or $\log \bar\mu$), either because of the cancellation between near/far zone contributions we discussed above, or because they are absorbed into coefficients which can be removed by field-redefinitions. The left-over factors of $\log \omega r$ from the radiation region are also omitted, except for the surviving term shown in \eqref{QQtail}. This is the only contribution which is not proportional to a quantity that vanishes on-shell.\vskip 4pt After all is said and done, the resulting effective Lagrangian can be written as \beq \cL_{\rm 3,4PN} =\cL_{\rm 0123PN}+\cL_{\rm 4PN}\,.\eeq  The $\cL_{\rm 0123PN}$ is the finite part of the Lagrangian in harmonic gauge to 3PN order, as shown in \cite{Foffa:2019rdf} (see also \cite{Foffa:2011ub}), except for the  $G^3$ and $G^4$ contributions, which for us here should read
\bea
\label{eq:3PN}
\cL_{\rm 3PN}^{G^{3,4}}&=&\frac{G^3 m_1^3 m_2}{r^3}\left[\frac{209}{18}\bv_1^2
-\frac{118}{9}\bv_1\cdot \bv_2+\frac{5}{4}\bv_2^2-\frac{355}{12}(\bv_1\cdot \bn)^2+\frac{82}{3}(\bv_1\cdot \bn) (\bv_2\cdot \bn)+\frac{3}{2}(\bv_2\cdot \bn)^2\right]\nonumber\\
&+&\frac{G^3 m_1^2 m_2^2}{r^3}\left[-\frac{305}{72}\bv_1^2+\frac{439}{144}\bv_1\cdot \bv_2+ \frac{383}{24}(\bv_1\cdot \bn)^2-\frac{889}{48}(\bv_1\cdot \bn)(\bv_2\cdot \bn)\right. \nn \\ &+& \left.\frac{41\pi^2}{64}\left( \bv\cdot \bv_1-3(\bv\cdot \bn) (\bv_1\cdot \bn)\right)\right]-\frac{3}{8}\frac{G^4 m_1^4 m_2}{r^4}-\frac{67}{3}\frac{G^4 m_1^3 m_2^2}{r^4} + \left(1 \leftrightarrow 2 \right)\,,
\eea
where we introduced $\bn \equiv \br/r$. The modified version of the 3PN Lagrangian is due to undoing integration by parts  in the terms shown in \cite{Foffa:2019rdf}, which was required here to properly identify the coefficients of the logarithms in the near zone.\footnote{More concretely, the difference is in the term $\tfrac{1}{r^3} (\br\cdot \ba_1) \log \mu r$ we used earlier, instead of the expression $\tfrac{1}{r^3}(\bv\cdot \bv_1- \frac{3}{r^2} (\br\cdot \bv)(\br\cdot \bv_1))\log \mu r $, which appears in the literature.}\vskip 4pt For the 4PN effective action, it can be written as follows:
\beq
\label{total4PN}
\cL_{\rm 4PN} =\frac 7{256}m_1v_1^{10}+\cL_{\rm 4PN}^G+\cL_{\rm 4PN}^{G^2}+\cL_{\rm 4PN}^{G^3}+
\cL_{\rm 4PN}^{G^4}+\cL_{\rm 4PN}^{G^5} + \cL_{\rm 4PN}^{G^2 \log v}\,,
\eeq
where the last term, introduced in \eqref{G2nonloc}, encodes the nonlocal-in-time contribution from the far zone due to the tail effect. We have absorbed the celebrated $\frac{41}{30}$ in \eqref{eq:RRnl} into the local part of the effective action, and subsequently reduced it using the double-zero trick, turning it into an ${\cal O}(G^4)$ contribution.  This means that, in practice, the first two local terms from the near zone, $\cL_{\rm 4PN}^G+\cL_{\rm 4PN}^{G^2}$, remain as reported in \cite{Foffa:2012rn}~(see~Eqs.~(13)~and~(26)), while the others take the form: 
\bea\label{4PNG3reg}
\cL_{\rm 4PN}^{G^3}&=&\frac{G^3 m_1^3 m_2}{r^3}\left[\ba_1\cdot\bv_2 \left(\frac{3763}{240}\bv_2\cdot \br
-\frac{18719}{720}\bv_1\cdot \br \right)\right. \\ &+& \left.  \ba_1\cdot \br \left(-\frac{18719}{1440}\bv_1^2-\frac{95119}{7200}\bv^2+\frac{1309}{48}(\bv_2\cdot \bn)^2-\frac{75}{4}(\bv_1\cdot \bn)(\bv_2\cdot \bn) \right)\right.\nonumber\\
&+&\frac{3763}{480} (\ba_2\cdot\br) \bv_1^2-\frac{231}{160}\bv_1^4+\frac{1397}{480}\bv_1^2 \bv_2^2 -\frac{433}{60}\bv_1^2 (\bv_1\cdot \bv_2)+\frac{43}{2}(\bv_1\cdot \bv_2)( \bv\cdot \bv_2)+\frac{91}{16}\bv_2^4\nonumber\\
&+&\bv_1^2\left(\frac{3463}{160}({\bv_1\cdot \bn})^2-\frac{1047}{20}(\bv_1\cdot \bn)( \bv_2\cdot \bn)
+\frac{3923}{160}({\bv_2\cdot \bn})^2\right)\nn \\ &+& \left.\bv_1\cdot \bv_2\left(7(\bv_1\cdot \bn)( \bv_2\cdot \bn)+\frac{43}{16} ({\bv_1\cdot \bn})^2-2({\bv_2\cdot \bn})^2\right)\right. \nn \\ &+& \left. \bv_2^2\left(\frac{7}{4}({\bv_2\cdot \bn})^2-\frac{1}{8}({\bv_1\cdot \bn})^2-\frac{7}{2}(\bv_1\cdot \bn)( \bv_2\cdot \bn)\right)\right. \nn \\ &+& \left. ({\bv_1\cdot \bn})^2\!\left(\frac{5}{2}({\bv_1\cdot \bn})^2+\frac{35}{16}(\bv_1\cdot \bn )(\bv_2\cdot \bn)-\frac{15}{4}({\bv_2\cdot \bn})^2\right) \!\right]\nonumber\\
&+&\frac{G^3 m_1^2 m_2^2}{r^3}\left[\ba_1\cdot \br \left(\left(\frac{349207}{7200}-\frac{43}{128}\pi^2\right)\bv^2+\left(\frac{123\pi^2}{128}-\frac{2005}{96}\right)({\bv_2\cdot \bn})^2\right)\!\!\right.\nonumber\\
&+&  \left( (\ba_1\cdot \br )\bv_1^2+2(\bv_1\cdot \bn) (\bv_2\cdot \ba_1)\right)\left(\frac{1099}{288}-\frac{41\pi^2}{128}\right)+\frac{383}{192}\bv_1^4 \nn \\ &+&  \left(\frac{21427}{480}+\frac{133\pi^2}{1024}\right)\left(\bv_1^2 \bv^2-2(\bv_1\cdot \bv_2) (\bv\cdot \bv_1)\right)-\frac{55}{24}\bv_1^2 (\bv_1\cdot \bv_2)\nonumber\\
&+& \bv_1\cdot \bv_2\left(\frac{32887}{150}({\bv_1\cdot \bn})^2- \frac{33487}{150} (\bv_1\cdot \bn)( \bv_2\cdot \bn)-\frac{447\pi^2}{256}(\bv\cdot \bn)(\bv_1\cdot \bn)\right)\nonumber\\
&+& \bv_1^2\left(\frac{270521}{1200}(\bv_1\cdot \bn)( \bv_2\cdot \bn)-\frac{275321}{2400}({\bv_1\cdot \bn})^2-\frac{64799}{600}({\bv_2\cdot \bn})^2+\frac{447\pi^2}{512}({\bv\cdot \bn})^2\right)\nonumber\\
&+&\left. ({\bv_1\cdot \bn})^2 \left(\frac{155947}{2880}({\bv_1\cdot \bn})^2-\frac{155977}{720}(\bv_1\cdot \bn)( \bv_2\cdot \bn)+\frac{78911}{480}({\bv_2\cdot \bn})^2\right. \right. \nn \\ &-& \left. \left. \frac{2155\pi^2}{1024}\left(({\bv_1\cdot \bn})^2-4(\bv_1\cdot \bn)( \bv_2\cdot \bn)+3({\bv_2\cdot \bn})^2\right)\right) \right] +\left(1\leftrightarrow 2\right)\,, \nn\\
\label{4PNG4reg}
\cL_{\rm 4PN}^{G^4}&=&\frac{G^4 m_1^4 m_2}{r^4}\left[-\frac{98549}{3600}\bv_1^2+\frac{95849}{3600}\bv_1\cdot \bv_2+\frac{15}{16}\bv_2^2+\frac{103949}{900}({\bv_1\cdot \bn})^2\right.  \\ &-& \left. \frac{105299}{900}(\bv_1\cdot \bn)( \bv_2\cdot \bn)+\frac{9}{4}({\bv_2\cdot \bn})^2\right] \nn \\ &+& \frac{G^4 m_1^3 m_2^2}{r^4}\left[-\left(\frac{104569}{7200}+\frac{15}{32}\pi^2\right)\bv_1^2 + \left(\frac{103}{16}\pi^2-\frac{11923}{240}\right)\bv_1\cdot \bv_2+\left(\frac{542659}{7200}-\frac{191}{32}\pi^2\right)\bv_2^2\right. \nonumber\\
&+& \left(\frac{659}{96}\pi^2-\frac{125209}{7200}\right)({\bv_1\cdot \bn})^2+\left(\frac{296893}{720}-\frac{1715}{48}\pi^2\right)(\bv_1\cdot \bn )(\bv_2\cdot \bn)\nn \\ &+& \left. \left(\frac{2771}{96}\pi^2-\frac{871207}{2400}\right)({\bv_2\cdot \bn})^2\right]+\left(1\leftrightarrow 2\right)\,,\nonumber\\
\label{4PNG5reg}
\cL_{\rm 4PN}^{G^5}&=&\frac{3}{8}\frac{G^5 m_1^5 m_2}{r^5}+\frac{G^5 m_1^4 m_2^2}{r^5}\left(\frac{94931}{3600}+\frac{105}{32}\pi^2\right)+\frac{G^5 m_1^3 m_2^3}{r^5}\left(\frac{225839}{2400}-\frac{71}{32}\pi^2\right)+\left(1\leftrightarrow 2\right)\,.\nonumber\\
\eea
Our final result can be shown to be equivalent to the one obtained in \cite{Damour:2014jta,Marchand:2017pir} leading, for instance, to the same expression for the binding energy and periastron advance.

\subsubsection*{ Lorentz Invariance}

There is only one minor caveat in the steps leading to the above expression for the renormalized Lagrangian. Prior to the removing the UV $\log \bar\mu r$'s, the effective theory is Lorentz invariant to 4PN order (up to double-zeros), with $\mu$ treated as a constant. Since $ r= |\bx_1-\bx_2|$ transforms under a boost \cite{boost}, the choice $\bar\mu=1/r$, which removes the logarithms, has the result of making the action Lorentz invariant only on-shell. Of course, this is inconsequential, as any choice of $\mu$ is physically equivalent. There is, nonetheless, a simple way to recover manifest Lorentz invariance of the effective action, by means of a worldline redefinition. In particular, it is easy to see that the transformation
$\delta \bx_a(t)= -11 G^2 m_a^2(\bv_b\cdot \br)^2/(3r^2)\, \ba_a$ introduces an extra term,
\beq
\label{eq:DLLI}
\Delta \cL_{\rm LI} = \frac{11 G^2 m_1^3}{3}\frac{(\bv_2\cdot \br)^2}{r^2} \left(\ba_1+\frac{G m_2 \br }{r^3}  \right)\cdot \ba_1  + \left(1\leftrightarrow 2\right)\,,
\eeq
which readily reinstates manifest (perturbative) Lorentz invariance. Notice, as expected, that it vanishes on-shell. While it is straightforward to introduce this or other similar terms, we did not include it since it does not affect observable quantities.

\section{Discussion}\label{sec:disc}

The existence of intermediate IR/UV divergences, as well as the explicit cancellation, is entirely a byproduct of the split into near and far zone, the asymptotic expansion of Feynman integrals \cite{Beneke:1997zp} and the use of the method of regions with potential and radiation modes, e.g. \cite{Rothstein:2003mp,Manohar:2018aog}. In principle, we do not need to perform this splitting, since we can always work with a point-particle effective theory and relativistic propagators, performing instead the Post-Minkowskian (PM) expansion.\footnote{The computation of the classical limit of scattering amplitudes in the PM framework has recently received renewed attention, due to their ability to extract the conservative Hamiltonian of the two-body problem at a given PM order through a matching calculation \cite{ira1,ira2,zvi1} (see also \cite{Holstein:2008sx,Galley:2013eba,Vaidya:2014kza,Damour:2017zjx,Guevara:2018wpp,Laddha:2018vbn,Kosower:2018adc,Caron-Huot:2018ape,Vines:2018gqi,Addazi:2019mjh,Antonelli:2019ytb,Henn:2019rgj} and references therein).} It is only for convenience, and to separate the relevant physics in the near and far regions, that potential and radiation modes are introduced. The price to pay, however, for the simplification of the Feynman integrals, is the introduction of new divergences which are not present in the original theory. In our case, spurious IR/UV singularities appear from the near/far expansions of the iterated Green's functions in PN, already at ${\cal O}(G^2)$ (i.e. `one-loop').\vskip 4pt For example, let us return to the topology in Fig.~\ref{fig:Y}, and consider the full propagators without expanding in the regions of the non-relativistic limit. Concentrating on the IR properties, it is easy to see then that the diagram is dominated by a triangle integral,
\beq
\label{eq:Dp}
\Delta(p,\sigma_1) = \int d\sigma_1^\prime \int \frac{d^D k}{(2\pi)^D} \frac{1}{k^2(p-k)^2} e^{i k_\mu(x_1^\mu(\sigma_1)-x_1^\mu(\sigma_1^\prime))}\,,
\eeq 
in $D$ dimensions. Notice that for $D=4$ this integral is both UV and IR finite for bound states. The remaining integral consists of a Fourier transform. The exact form depends on the numerator, which carries extra powers of momenta. Schematically, it takes the form
\beq
\int d\sigma_1 d \sigma_2 \int \frac{d^D p}{(2\pi)^D} \Delta(p,\sigma_1) \frac{p^\alpha p^\beta}{p^2} e^{i p_\mu \left(x^\mu_1(\sigma_1)-x^\mu_2(\sigma_2)\right)}\,.\label{fulle}
\eeq
In the static limit, the triangle in \eqref{eq:Dp} collapses into a $d=(D-1)$ integral, such that we end up with a three-dimensional Fourier transform to obtain the $G^2$ correction to the potential at 1PN, proportional to $\int d^3\bp\, |\bp|^{-1} e^{i\bp\cdot \br} \propto 1/r^2$ in $d=3$, as expected. For the general case the integral is more complicated, however, the result is IR finite by power counting.\vskip 4pt

The situation changes when we use the method of regions \cite{Rothstein:2003mp,Manohar:2018aog}. Expanding the propagators with potential modes in powers of $p_0/|\bp|$ inside the integrals, it is easy to see one of the terms is of the form
\beq
\ba_1^i \ba_2^j \int dt \int \frac{d^3 \bp}{(2\pi)^3} \frac{\bp^i \bp^j}{|\bp|^5} e^{-i\bp\cdot\left(\bx_1(t)-\bx_2(t)\right)}\,,
\eeq 
which is logarithmically IR divergent. The key difference is that, for quasi-instantaneous interactions, the $p_0$ integrals are traded for time derivatives on the worldlines, which in the classical EFT are treated as external (non-propagating) sources. At the same time, the IR properties of the diagram change due to the $1/|\bp|$ factors. These IR divergences, which do not belong to the near zone, are therefore clearly an artifact of the expansion into regions. It comes as no surprise then that they will be linked to (and cancel against) singularities associated to radiation modes.\vskip 4pt 

For the conservative contribution from the far zone, we must consider on-shell modes with $p_0 \simeq |\bp| \simeq 1/\lambda_{\rm rad}$, where $\lambda_{\rm rad} \sim r/v$, as well as long-distance quasi-instantaneous (potential) modes with $p_0 \ll |\bp| \sim 1/\lambda_{\rm rad}$. The latter build up the (Kerr) background geometry produced by the binary as a whole in the far zone. The propagators for the (off-shell) potential modes are expanded as in the near region, but not the (on-shell) radiation modes. In the non-relativistic limit, we have $\bp\cdot\br \simeq r/\lambda_{\rm rad} \ll 1$ and therefore the (spatial part of the) exponential in \eqref{fulle} is expanded instead. Hence, rather than having a one-loop integral plus a Fourier transform for the topology in Fig.~\ref{fig:Y}, with radiation modes we end up with a two-loop type integral. In the EFT approach this is represented by a diagram with the topology of Fig.~\ref{nltail}, with the binary described as a localized source endowed with a series of multipole moments.\footnote{For more complicated topologies, as in Fig.~\ref{fig:3pnUV}, the contribution from radiation modes is also described by a diagram similar to Fig.~\ref{nltail}, but with the source multipoles incorporating higher order correction in $G$. This is handled in the EFT approach in two steps. First, by matching the source multipoles in the long-distance EFT to the worldline effective theory (integrating out the potential modes for the two bodies). Subsequently, by computing the (two-loop) tail diagram. See  Appendix~\ref{sec:appA} for more details.} For the case at hand in \eqref{fulle}, after expanding in the radiation region, we find first the term
\beq
\int \frac{d\omega}{\pi} \int  \frac{d^3 \bp}{(2\pi)^3}  \frac{d^3 \bk}{(2\pi)^3} \, \bx^i_1(\omega) \bx^j_1(\omega) \frac{\bk^i\bk^j}{\bp^2 k^2(p+k)^2}\,,
\eeq
with $k^2\equiv \omega^2-\bk^2+i\epsilon$ and $(p+k)^2 = \omega^2-(\bq+\bk)^2+i\epsilon$. This (two-loop) integral is UV divergent, and proportional to derivatives of the dipole moment associated with the binary. Adding extra terms in the expansion of the exponential it is straightforward to show the far zone contribution is UV divergent, and can be written in terms of derivatives of the source multipoles. As for the IR poles from the potential region, the UV divergence is likewise artificial, and it is due to performing a multipole expansion. As anticipated, after the subtraction of the zero-bin, by construction the poles cancel out against the equally unphysical divergence from the near region \cite{Porto:2017dgs,Porto:2017shd}. Because the binary is shrunk to a point, some of the divergences introduced by the multipole expansion will match into self-energy terms in the potential region. That is the reason additional scaleless integrals are needed to make manifest the cancellation. In the above case, the dipole term does not contribute to physical quantities. However, starting with the (trace-free) quadrupole, radiation modes contribute to the conservative sector, as explained~here.\vskip 4pt 

Needless to say, as a systematic approximation to the full answer, the PN expansion of the relativistic expression must be captured by the decomposition into potential and radiation modes, albeit introducing intermediate spurious poles. Yet,
to correctly perform the expansion into regions, and in particular when different regions overlap as it happens in our case, we had to implement the zero-bin subtraction \cite{zb}. In this procedure, in which the IR-sensitivity in the near zone is removed from the region of integration, there are left-over UV poles. The latter cancel out against the UV poles produced by the multipole-expanded computation involving radiation modes. Had we not expanded the integral(s) into regions, these spurious divergences would not appear. Yet, their existence is not a mystery in the EFT framework. The method of regions is a useful tool to isolate the relevant physics one scale at a time, but the procedure may introduce artificial divergences which must be properly removed \cite{Porto:2017dgs,Porto:2017shd}.\footnote{As a perk of using the method of regions, the appearance of UV poles in the long-distance theory allows us to resum (universal) logarithmic corrections using the renormalization group \cite{Galley:2015kus,Porto:2017dgs,Porto:2017shd}.}\vskip 8pt

In conclusion, in this paper we have derived the renormalized effective action describing the dynamics of binary systems to 4PN order. We have shown how to implement dim.~reg. when IR and UV divergences appear in intermediate steps, paying especial attention to the contribution from scaleless integrals. We have  renormalized away the near zone UV poles by the use of counter-terms in the point-particle effective theory.  As expected from the effacement theorem \cite{damour}, the counter-terms can be removed by field-redefinitions at this order, e.g. \cite{nrgr}.\footnote{\label{footspin} Let us stress an important point. In order to remove the $c_{a\dot v}$ coefficient by a field-redefinition, spin and finite-size effects must be ignored. Once spin is included, the field-redefinition $\delta x^\mu = - c_{a\dot v} \dot v^\mu$ leads~to \beq 
-c_{a\dot v} \left(a_\mu - \frac{1}{2m} R_{\mu\nu\alpha\beta} v^\nu S^{\alpha\beta} + \cdots \right)\dot v^\mu\,,\eeq
with the ellipses including terms quadratic in the spin \cite{review}. Hence, the structure of UV poles must be such that they are canceled by the above operator (plausibly including other spin-dependent counter-terms), with the precise coefficient given in \eqref{cadv}. This provides a non-trivial consistency check for our procedure.} We have also shown how the spurious IR divergences in the potential region are handled by means of the zero-bin subtraction \cite{zb}, which unambiguously removes from the Feynman diagrams the region of integration which does not belong to the near zone \cite{Porto:2017dgs,Porto:2017shd}. The left-over UV poles which result from this procedure cancel out against equally unphysical divergences in radiation-reaction~effects.\vskip 4pt Our derivation of the 4PN renormalized Lagrangian differs from the methodologies used in \cite{Jaranowski:2013lca,Bini,Damour:2014jta,Damour:2016abl,Jaranowski:2015lha,Bernard:2015njp,Bernard:2016wrg,Bernard:2017bvn,Marchand:2017pir}, mainly for two reasons. First of all, we have implemented a regularization-independent procedure to remove IR divergences which uniquely determines the gravitational potential, thus not requiring the introduction of ambiguity-parameters nor information outside of the realm of the PN expansion. Secondly, contrary to what is ultimately the case in the Fokker-action approach \cite{Marchand:2017pir}, we do not remove near zone IR poles combined with UV divergences in tail terms by means of redefinitions of the particles' worldlines.\footnote{Let us emphasize that the lack of a systematic treatment of IR/UV divergences can potentially lead to inconsistencies at higher PN orders, in particular when UV divergences in the potential region (and related length scales and logarithms) are associated to physical finite-size effects. Unlike wordline redefinitions, counter-terms and renormalized parameters in the effective theory can still be applied in such case, with operators that do not vanish on-shell.}
Instead, we have identified the IR/UV singularities from the different regions and demonstrated the explicit cancellation of the spurious near/far zone divergences after the subtraction of the zero-bin. While the procedure is independent of the regulator, we have performed all of the calculations within the confines of dim. reg.,
as it is customary in an EFT approach. This is in contrast to the results in \cite{Jaranowski:2013lca,Bini, Damour:2014jta,Damour:2016abl,Jaranowski:2015lha,Bernard:2015njp,Bernard:2016wrg,Bernard:2017bvn,Marchand:2017pir} where an additional regulator appears to be a necessity. Hence, in comparison with the Fokker-action \cite{Blanchet:2013haa} and ADM \cite{Schafer:2018kuf} approaches, in our opinion the EFT formalism provides a more systematic derivation of the conservative dynamics, which furthermore can be naturally extended to all PN orders without ambiguity-parameters nor extra regulators.\vskip 4pt

We have demonstrated the ability of the EFT approach to tackle intricate calculations in a systematic and scalable fashion within the PN framework. Yet, the future enterprise in precision GW physics is vast, with the associated complexity increasing in every iteration. In order to move forward, reaching the physically motivated threshold at 5PN order (and beyond) where the first finite-size operators appear in \eqref{eq:spp} \cite{Porto:2016zng, Porto:2017lrn}, may require significant advances both in our understanding of the theoretical foundations and computational efficiency (see \cite{Foffa:2019hrb} for partial results at 5PN in the EFT approach, also \cite{Blumlein:2019zku}). In a  parallel development, the study of scattering amplitudes and other ideas from particle physics have opened up new routes to simplify calculations for the conservative sector, and plausibly also for the radiated power, e.g. \cite{ira1,ira2,zvi1, Holstein:2008sx,Galley:2013eba, Vaidya:2014kza,Damour:2017zjx,Guevara:2018wpp,Laddha:2018vbn,Kosower:2018adc,Caron-Huot:2018ape,Vines:2018gqi,Addazi:2019mjh,Antonelli:2019ytb,Henn:2019rgj}. In principle, these new tools can further streamline the relevant PN computations within~the~EFT~approach, which we find is a natural venue for future explorations.

 \subsubsection*{Acknowledgements} 

We thank Luc Blanchet, Thibault Damour, Guillaume Faye, Adam Leibovich and Gerhard Sch\"afer for very useful discussions and comments. We would like to thank ICTP-SAIFR as well as the Mainz Institute for Theoretical Physics, for the support to organize the workshops ``Analytic Methods in General Relativity,"\footnote{\url{http://www.ictp-saifr.org/gr2016}} and ``The Sound of Spacetime: The Dawn of Gravitational Wave Science,"\footnote{\url{https://indico.mitp.uni-mainz.de/event/124/}} respectively, where this work originated and preliminary results were presented. We~thank Nordita and the organizers of the workshop ``QCD Meets Gravity IV",\footnote{\url{http://www.nordita.org/qcd2018}} for hospitality while this work was being completed. S.F. is supported by the Fonds National Suisse and by the SwissMap NCCR. 
R.A.P.  acknowledges financial support from the ERC Consolidator Grant ``Precision Gravity: From the LHC to LISA"  provided by the European Research Council (ERC) under the European Union's H2020 research and innovation programme (grant agreement No. 817791), as well as from the Deutsche Forschungsgemeinschaft (DFG, German Research Foundation) under Germany's Excellence Strategy (EXC 2121) `Quantum Universe' (390833306). R.A.P. would like to thank also the Simons Foundation and FAPESP (Young Investigator Awards) for support during the early stages of this work, and the organizers and participants of the `Simons Foundation Symposium: Amplitudes meet Cosmology'\footnote{\url{https://www.simonsfoundation.org/event/amplitudes-meet-cosmology-2019/}} for the opportunity to present this work and for helpful discussions. R.S. has been supported for part of the duration of the present work by the FAPESP grant 2012/14132-3, and wishes to thank the Physics Department  at the University
of Geneva for hospitality and support during his visits. R.S. acknowledges the High-Performance Computing Center at UFRN.

 \appendix
 
\section{Feynman rules \& Master integrals}\label{sec:app0}
To compute the contribution from the near zone it is convenient to decompose the metric field ($g_{\mu\nu}$) in terms of scalar, vector and tensor perturbations: $(\phi$, $\bA$, $\gamma_{ij})$. The gauge-fixed Einstein-Hilbert action in $(d+1)$-dimensions, 
\be
S_{\rm EH}=-\frac1{16 \pi G_d}\int d^{d+1}x\sqrt{-g}\pa{R[g]-\frac 12\Gamma^\mu\Gamma_\mu}\,,
\ee
then reads, in harmonic gauge $\Gamma^\mu \equiv \Gamma^{\mu}_{\alpha\beta}g^{\alpha\beta}$,
\renewcommand{\arraystretch}{1.4}
\be
\label{eq:sEH_KK}
\bba{rcl}
\displaystyle S_{\rm EH}&=&\displaystyle \int {\rm d}^{d+1}x\sqrt{-\gamma}
\left\{\frac{1}{4}\left[(\nabla\sigma)^2-2({\nabla}\sigma_{ij})^2-\left(\dot{\sigma}^2-2(\dot{\sigma}_{ij})^2\right){\rm e}^{\frac{-c_d \phi}{\Lambda}}\right]- c_d \left[({\nabla}\phi)^2-\dot{\phi}^2 {\rm e}^{-\frac{c_d\phi}{\Lambda}}\right]\right.\\
&&\displaystyle
+\left[\frac{1}{2} F_{ij}F^{ij}+\left(\nabla \cdot \bA\right)^2 -{\dot{\bA}}^2 {\rm e}^{-\frac{c_d\phi}{\Lambda}} \right]
{\rm e}^{\frac{c_d \phi}{\Lambda}}+\frac 2\Lambda\paq{\pa{F_{ij}\bA^i\dot{\bA^j}+{\bA}\cdot \dot{\bA}(\nabla\cdot \bA)}
{\rm e}^{\frac{c_d \phi}{\Lambda}}-c_d\dot{\phi}{\bA}\cdot \nabla\phi}
\\
&&\displaystyle
+2 c_d \left(\dot{\phi}\nabla\cdot{\bA}-\dot{\bA}\cdot\nabla\phi\right)
+\frac{\dot{\sigma}_{ij}}{\Lambda}\left(-\delta^{ij}A_l\hat{\Gamma}^l_{kk}+ 2A_k\hat{\Gamma}^k_{ij}-2A^i\hat{\Gamma}^j_{kk}\right)-c_d\frac{\dot{\phi}^2{\bA}^2}{\Lambda^2}\\
&&\displaystyle
-\left.\frac{1}{\Lambda}\left(\frac{\sigma}{2}\delta^{ij}-\sigma^{ij}\right)
\left({\sigma_{ik}}^{,l}{\sigma_{jl}}^{,k}-{\sigma_{ik}}^{,k}{\sigma_{jl}}^{,l}+\sigma_{,i}{\sigma_{jk}}^{,k}-\sigma_{ik,j}\sigma^{,k}
\right)\right\}+\cdots\,,
\ea
\ee
where we used $\Lambda\equiv 1/\sqrt{32 \pi G_d}$, $\sigma_{ij} \equiv  \gamma_{ij}-\delta_{ij}$ and $\sigma\equiv \sigma_{ii}$. The above expression, along with the expansion of $S_{\rm pp}[x^\alpha_a(\tau_a)]$ in \eqref{eq:spp}, allows us to derive all of the Feynman rules. For instance the propagators for each field take the form (in mixed direct-Fourier space):
\be
\label{eq:props}
 \frac{1}{2} P^{aa}\delta_{ab}
(2\pi)^{d}\delta^{(d)}(\bp+\bq){\cal P}(\bp^2,t_1,t_2)\delta(t_1-t_2)\,,
\ee
where $P^{\phi\phi}=-\frac{1}{c_d}$, $P^{A_iA_j}=\delta_{ij}$, 
$P^{\sigma_{ij}\sigma_{kl}}=-\left(\delta_{ik}\delta_{jl}+\delta_{il}\delta_{jk}+(2-c_d)\delta_{ij}\delta_{kl}\right)$ 
and
\be
\label{eq:props_fac}
{\cal P}(p^2,t_1,t_2)=\frac{i}{\bp^2-\partial_{t_1}\partial_{t_2}}\,.\ee
In the non-relativistic limit the time derivatives are expanded in Taylor series, and ultimately applied to the external (world-line) sources. As discussed in sec.~\ref{sec:disc}, this is the reason for the spurious near-zone IR divergencies.\vskip 4pt

All integrals required for the evaluation of the 4PN effective action can be reduced to combinations of the following master integrals (see \cite{Foffa:2012rn, Foffa:2019rdf} for details):
\be
\label{fourier}
\int \frac{d^dp}{(2\pi)^d}\frac{e^{i{\bp}\cdot{\br}}}{\bp^{2a}}=
2^{-2a}\pi^{-d/2}\frac{\Gamma(d/2-a)}{\Gamma(a)}\, r^{2a-d}\,,
\ee
\be
\label{oneloop}
\int \frac{d^dk}{(2\pi)^d}\frac 1{\pa{\bk-\bp}^{2a}\bk^{2b}}=\frac 1{(4\pi)^{d/2}}
\frac{\Gamma(d/2-a)\Gamma(d/2-b)\Gamma(a+b-d/2)}{\Gamma(a)\Gamma(b)\Gamma(d-a-b)}\,,
\ee
with the exception of some contributions at ${\cal O}(G^5)$, which
can be mapped into four-loop (massless) two-point functions.
The integrals involved in these diagrams have been computed in terms
of more complex master integrals (see \cite{Foffa:2016rgu} for a detailed discussion, also \cite{Damour:2017ced}). Notice that eq.~\eqref{fourier} can only
have IR poles (for $a=\frac d2+n$, with integer $n\geq 0$),
while eq.~\eqref{oneloop} may contains both IR and UV poles
(e.g. the latter occurring when $a+b=\frac d2-n$).

\section{Spurious IR/UV poles}\label{sec:appA} 

The EFT for the radiation region in \eqref{eq:srad}, augmented by the terms in \eqref{eq:multeqm}, leads to several UV divergences in the conservative sector of the radiation-reaction force. After inserting the different (source) couplings into tail-type diagrams, similar to Fig.~\ref{nltail}, we find:
\bea
\cL_{\phi^3,\, \rm 4PN}^{\rm UV\, (far)}&=&\frac{G^2}{\epsilon_{\rm UV}}
\left[\Big(M+T^{kk}\Big)\left(-\frac 1{60}\left(2\left({I_0^{ij(3)}}\right)^2+\left({I_0^{ii(3)}}\right)^2\right)-\left(\dot M+T^{kk(1)}+\frac 13 I_0^{kk(3)}\right)\left(\dot M+ T^{ii(1)}\right)\right.\right.\nonumber\\ &&\left.\left.-\frac13\left(\ddot \bX^{i}+ \ddot \bT^i\right)^2
-\frac{1}{15} I_0^{ikk (4)}\left(\ddot \bX^{i}+ \ddot \bT^i\right)\right) -\frac{2}{3}(\dot M+T^{kk(1)}) \left(\ddot \bX^{i}+ \ddot{{ \bT}}^i\right)\left(\dot \bX^{i}+ \dot{{ \bT}}^i\right)+\right. \nn \\ && \left. \frac{1}{30}\left({I_0^{kk(2)}}\delta^{ij}+2I_0^{ij(2)}\right)\left(\ddot{\bX}^i\ddot{\bX}^j+2 \dot{\bX}^i \dddot{\bX}^j\right)\right]\,,\\
\cL_{A^2\phi, \, \rm 4PN}^{\rm UV\, (far)}&=&\frac{G^2}{\epsilon_{\rm UV}}\left[(M+T^{kk})\left(4 \dot{{\bP}}^2+\frac 43  \left(M^{ij(2)}\right)^2+\frac 43\dot \bP^i M^{ikk(3)}\right) \right. \nn \\ && \left. + \frac23 \dot{{\bP}}^2I_0^{kk(2)}+\frac83 {M}^{ij(1)}\dot{\bP}^i\dot{\bX}^j\right]\,,\\
\label{eq:sigma2phi}\cL_{\sigma^2\phi\,\, \rm 4PN}^{\rm UV\, (far)}&=&\frac{2G^2M}{\epsilon_{\rm UV}}\left[ \left(T^{kk(1)}\right)^2- \left(T^{ij(1)}\right)^2\right]\,,\\
\cL_{A\phi^2,\, \rm 4PN}^{\rm UV\,(far)}&=&\frac{G^2}{3\epsilon_{\rm UV}}\left[ 4 \ddot \bX^i\bP^i\left(\dot M+ T^{kk(1)}\right)- \frac{2}{5}{M}^{kk(1)}{\dot \bX}^i {\dddot\bX}^i-\frac{1}{5}I_0^{kk (2)}\left({\dot \bP}^i \ddot\bX^i+{\bP}^i\dddot\bX^i\right)\right.\nonumber\\
&&\left. -\frac{2}{5}\dot{M}_{ij}\left(\dot{\bX}^i\dddot{\bX}^j+\dot{\bX}^j\dddot{\bX}^i\right)-\frac{2}{5}I_0^{ij(2)}\left(\dot{\bP}^i\ddot{\bX}^j+\bP^i\dddot{\bX}^j\right)\right]\,,\\
\cL_{A^2\sigma,\,\rm 4PN}^{\rm UV \,(far)} &=&-\frac{8G^2}{3\epsilon_{\rm UV}}T^{kk}\,\dot \bP^i\dot \bP^i \,,\\
\cL_{\sigma\phi^2,\, \rm 4PN}^{\rm UV\, (far)}&=&\frac{2G^2}{15\epsilon_{\rm UV}}\left(-{T}^{ij}\ddot \bX^i\ddot \bX^j+2T^{kk}\ddot \bX^i\ddot \bX^i\right)\,,\\
\cL_{A^3,\,\rm 4PN}^{\rm UV\,(far)}&=&-\frac{8G^2}{3\epsilon_{\rm UV}}\left( M^{kk(1)}\dot{\bP}^i\dot\bP^i+2  M^{ij(1)} \dot \bP^i \dot\bP^j\right)\,,\\
\label{eq:pAs}\cL_{\phi A\sigma,\, \rm 4PN}^{\rm UV\,(far)}&=& 0\,,\\
\label{eq:ps2} \cL_{A\sigma^2,\,\rm 4PN}^{\rm UV\,(far)}&=& 0\,,\\
\label{eq:sg3} \cL_{\sigma^3,\, \rm 4PN}^{\rm UV\,(far)}&=& 0\,,
\eea
where we have also split the answer in terms of scalar, vector and tensor modes, and reported the results in standard variables such that the variation of the action acts as usual. For instance, the 2PN and 3PN contributions proportional to $\dot{M}^2$ and $\ddot{{\bX}}^2$ mentioned in the text arise from the first two equations. Moreover, the $\phi^3$, $\sigma^2 \phi$ and $ A^2\phi$ couplings account for the contribution from the quadrupole moment leading to \eqref{eq:RRnl} at 4PN.  However, other (yet unphysical) terms appear, involving the following moments of the pseudo stress-energy tensor, ${\cal T}^{\alpha\beta}$,
\begin{eqnarray}
\label{eq:momenta}
&& M\equiv\int \, d^3\bx\, {\cal T}^{00}= -\frac{G m_1m_2}r\left(1-\epsilon_{\rm UV}\log{\bar\mu r}\right) + \sum_a m_a\left(1+\frac{\bv^2_a}{2}\right) +\cdots\,, \\
&& \bX^i\equiv\int \, d^3\bx\, {\cal T}^{00}\bx^i= \sum_{a\neq b} m_a\left[1+\frac{\bv^2_a}2
-\frac{Gm_b}{2r}\left(1-\epsilon_{\rm UV}\log{\bar\mu r}\right)\right]\bx_a^i+\cdots\,, \\
&& \label{eq:PP} \bP^i\equiv\int \, d^3\bx\, {\cal T}^{0i}= \sum_{a\neq b} m_a\left[\left(1+\frac{\bv^2_a}{2}\right)\bv_a^i\right. \\ && \hspace {4cm} \left. -\frac{G m_b}{2r}\left(\bv_a^i+\frac{1}{r^2} \br\cdot \bv_a\,\br^i\right)\left(1-\epsilon_{\rm UV}\log{\bar\mu r}\right)\right]+\cdots\,,\nn \\
&& I_0^{ij}\equiv\int \, d^3\bx\, {\cal T} ^{00}\bx^i\bx^j= \sum_{a} m_a \bx_a^i \bx_a^j +\cdots\,,  \\  && I_0^{ijk}\equiv\int \, d^3\bx\, {\cal T}^{00}\bx^i\bx^j\bx^k= \sum_{a} m_a\bx_a^i \bx_a^j \bx_a^k+\cdots\,,\\
&& M^{ij}\equiv \int \, d^3\bx\, {\cal T}^{0i}\bx^j= \sum_{a} m_a \bv_a^i \bx_a^j+\cdots\,,  \\ && M^{ijk}\equiv\int \, d^3\bx\, {\cal T}^{0i}\bx^j\bx^k= \sum_{a} m_a \bv_a^i \bx_a^j \bx_a^k+\cdots\,,\\
&& T^{ij}\equiv \int \, d^3\bx\, {\cal T}^{ij}= \sum_a m_a \bv^i_a \bv_a^j -\frac{G m_1m_2}{r^3}\left(1-\epsilon_{\rm UV}\log{\bar\mu r}\right)\br^i \br^j +\cdots\,, \label{eq:Tijap}\\
&&  T^{ijk}\equiv \int \, d^3\bx\, {\cal T}^{ij}\bx^k = \sum_{a\neq b} m_a\left[\bv^i_a \bv_a^j-\frac{Gm_b}{2r^3}\br^i \br^j\left(1-\epsilon_{\rm UV}\log{\bar\mu r}\right)\right]\bx_a^k\,,\\
&&\bT^k \equiv T^{iik}  = \sum_{a\neq b} m_a\left[\bv^2_a-\frac{G m_b}{2r}\left(1-\epsilon_{\rm UV}\log{\bar\mu r}\right)\right]\bx_a^k+\cdots\,, 
\end{eqnarray}
\begin{figure}[t!]
\centerline{\scalebox{0.9}{\includegraphics{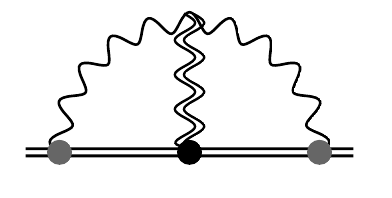}}}
\caption[1]{Diagram depicting a contribution due to the tail effect to the conservative sector of the radiation-reaction force coming from the $A^2\sigma$ coupling. Following \cite{Gilmore:2008gq}, we use a single-wavy line to represent the vector mode, $\bA$, while the double-wavy line accounts for the tensor, $\sigma$. The total binary's mass-energy, and moments of the pseudo stress-energy tensor, are described by the black and grey blob, respectively. All permutations of the fields must be considered.} \label{fig:sigma2far}
\end{figure}
and expanded to the desired order. The second equality is obtained by matching the stress tensor to the two-body effective action, including both potential and radiation modes, after the former are integrated out, see e.g. \cite{andirad,review}. Notice we have kept their values in $d$ dimensions, and included also the logarithmic terms. As we emphasized, even though the extra terms vanish on-shell, they are essential to remove all the unphysical poles in the intermediate computations. Adding all up, we arrive at the expressions given in \eqref{eq:divZB4pnG2}-\eqref{eq:divZB4pnG4}.\vskip 4pt It is instructive, however, to notice that the cancellation also occurs `polarization by polarization', or in other words, when the tensorial structure of the diagram is also taken into account. For example, to consider a specific case, let us look at the $A^2\sigma$ coupling. In the far zone, the relevant diagram(s) are summarized in Fig.~\ref{fig:sigma2far}. The result is given in \eqref{eq:sigma2phi}. Let us concentrate on the divergent parts, where we find 
\bea
\label{sigma2faruv}
\cL_{A^2\sigma\,\, \rm 4PN}^{\rm UV\, (far)} &=&  - \frac {8}{3\epsilon_{\rm UV}}G^2 m_1^2\left[m_2\left(2\bv_1^2(\ba_1\cdot \ba_2)+ \ba_1^2\bv_2^2\right) +m_1 \ba_1^2\bv_1^2\right] \\ 
&+&\frac {8}{3\epsilon_{\rm UV}}\frac{G^3 m_1^2m_2}{r}\left[ m_1 \ba_1^2+ m_2\, \ba_1\cdot \ba_2\right]+ (1\leftrightarrow 2)\,.\nn 
\eea
Notice it receives contributions both at ${\cal O}(G^2)$ and ${\cal O}(G^3)$, see \eqref{eq:PP} and \eqref{eq:Tijap}.\vskip 4pt On the other hand, the diagrams associated with the near zone computation are shown in Figs~\ref{fig:sig2phinear2} and \ref{fig:sig2phinear3}, for the $G^2$ and $G^3$ corrections, respectively (plus mirror images).
\begin{figure}[t!]
\centerline{\scalebox{0.85}{\includegraphics{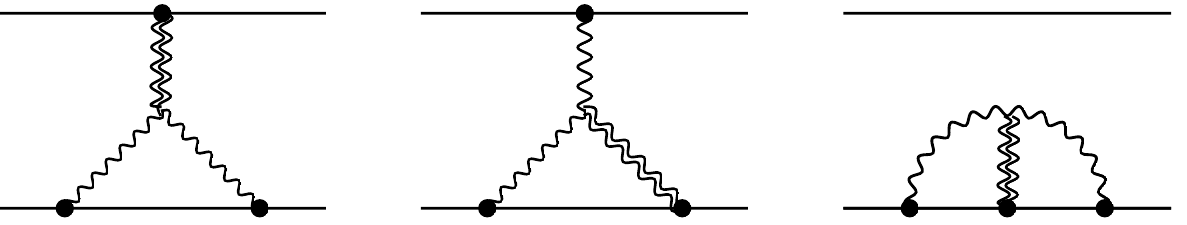}}}
\caption[1]{Near zone Feynman diagram at ${\cal O}(G^2)$ involving vector (single-wavy) and tensor (double-wavy) modes, associated with the far zone conservative contributions in Fig.~\ref{fig:sigma2far}, at the same order. Mirror images must be added.}
\label{fig:sig2phinear2}
\end{figure}
\begin{figure}[t!]
\centerline{\scalebox{0.85}{\includegraphics{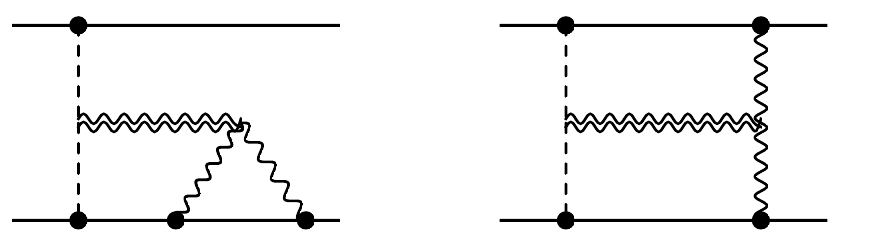}}}
\caption[1]{Near zone Feynman diagram at ${\cal O}(G^3)$ associated with the far zone conservative contributions in Fig.~\ref{fig:sigma2far}, at the same order. The dashed line represents the scalar mode. Mirror images must be added.}
\label{fig:sig2phinear3}
\end{figure}
The result for the $G^2$ topologies to 4PN order (which involve an expansion in $p_0/|\bp|$ for the {\it would-be} radiation modes) have both IR and UV divergences, given by  
\bea
\cL^{\rm IR/UV}_{\rm Fig.\,\ref{fig:sig2phinear2}}=\frac{8}{3} \frac{G^2 m_1^2 m_2}{\epsilon_{\rm IR}}\Big[ 2 \bv_1^2 (\ba_1\cdot\ba_2)+ \ba_1^2 \bv_2^2\Big]+ \frac{8}{3} G^2 m_1^3\left(\frac1{\epsilon_{\rm IR}}-\frac1{\epsilon_{\rm UV}}\right)\ba_1^2 \bv_1^2\,,\label{eq:fig7}
\eea
for the sum of the first two diagrams plus self-energy contribution, respectively. Recall that at ${\cal O}(G^2)$ only IR divergences are present, except for the scaleless integral. The UV pole from the latter, shown in \eqref{eq:fig7}, must be removed by the counter-terms, see sec.~\ref{sec:counter}. Whereas, after the zero-bin subtraction, the IR poles turn into UV divergences and exactly cancel the $G^2$ contribution~in~\eqref{sigma2faruv}, after adding the mirror diagrams.\vskip 4pt

The cancellation for topologies at ${\cal O}(G^3)$ is a bit more subtle at 4PN. The divergent parts of the relevant diagrams are given by
\bea
\cL^{\rm IR/UV}_{\rm Fig.\,\ref{fig:sig2phinear3}\,(\rm 1st)}&=& 
\frac{G^3 m_1^3 m_2}{\epsilon_{\rm UV}r^3}\left[
r\pa{-4 \bv_1^2 (\ba_1\cdot \bn)-2(\ba_2\cdot \bn){(\bv_1\cdot \bn)}^2+\frac 43(\ba_2\cdot\bv_1) (\bv_1\cdot \bn)}\right.\nonumber\\
&-&10 {(\bv_1\cdot \bn)}^2{(\bv_2\cdot \bn)}^2+8\bv_2^2{(\bv_1\cdot \bn)}^2-14\bv_1^2(\bv_1\cdot \bn)(\bv_2\cdot \bn) \nn\\ &+&\left.
8(\bv_1\cdot \bn)(\bv_2\cdot \bn) (\bv_1\cdot\bv_2)+\frac{136}5 \bv_1^2{(\bv_1\cdot \bn)}^2-2\bv_1^2\bv_2^2-\frac 43 (\bv_1\cdot\bv_2)^2\right. \nn \\ &+& \left. \frac{14}3\bv_1^2 \bv_1\cdot\bv_2
-\frac{136}{15}\bv_1^4\right]-\frac {8}{3\epsilon_{\rm IR}}\frac{G^3 m_1^3 m_2}r \ba_1\cdot \ba_1\,,\nonumber\\
\cL^{\rm IR/UV}_{\rm Fig.\,\ref{fig:sig2phinear3}\,(\rm 2nd)}&=&  =-\frac{8}{3\epsilon_{\rm IR}} \frac{G^3 m_1^2 m_2^2}{r} \,\ba_1\cdot \ba_2\,,
\eea
for the first and second diagram, respectively. Notice that, while the latter is IR divergent at 4PN, the former has both IR and UV poles. A similar diagram to the first graph in Fig.~\ref{fig:sig2phinear3}, with the $A$ and $\phi$ fields exchanged (not shown), is only UV divergent. For the other topologies at $G^3$, the first diagram shown in Fig.~\ref{fig:3pnUV} does not contribute for the given modes involved, neither does the second self-energy diagram in Fig.~\ref{fig:Y3}.  (Moreover, the first graph in Fig.~\ref{fig:3pnUV} is always UV divergent for every non-vanishing polarizations at 4PN.) There are no other contributions to 4PN~order. Once again, all of the UV poles are removed by counter-terms, while the IR singularity is handled by the zero-bin subtraction. As expected, the left-over UV pole cancels out against the associated contribution in \eqref{sigma2faruv} at ${\cal O}(G^3)$.\vskip 4pt  Notice that in the graphs at this order, the $\sigma$-propagator is attached to the scalar potential mode, rather than directly to the worldline. This is expected, since the radiation field couples also to the binding potential. In the far zone, this coupling is encoded in the expansion of the multipole moments in powers of Newton's constant. In our case, it is implicit in the ${\cal O}(G)$ corrections to the moment of the pseudo stress-energy tensor in \eqref{eq:Tijap}. (Something similar occurs when we match the multipole moments to compute the radiated power, e.g. \cite{andirad,srad1,srad2}.) We can show that, for the reasons discussed in sec.~\ref{sec:disc}, a similar cancellation of spurious divergences is at work for each one of the polarizations involved, order by order in $G$, to 4PN order.\footnote{It is straightforward to show, for instance, there is no contribution from the $A\sigma^2, \phi A\sigma$ and $\sigma^3$ couplings at 4PN order from either side, i.e. \eqref{eq:pAs} - \eqref{eq:sg3}.}
 
\phantomsection
\addcontentsline{toc}{section}{References}
\bibliographystyle{utphys}
\bibliography{4pnRefs}

\end{document}